\documentclass[12pt]{article}
\usepackage{epsfig}

\def\nn{\nonumber}
\def\im{\mbox{ Im }}
\def\ie{{\em i.e.}}

\def\eg{{\em e.g.}}
\def\tr{\mbox{ Tr }}

\begin{document}

\title{NUCLEON QCD SUM RULES IN NUCLEAR MATTER
INCLUDING
FOUR-QUARK CONDENSATES}
\author{
E.G. Drukarev, M.G. Ryskin, V.A. Sadovnikova\\
Petersburg Nuclear Physics Institute, \\
Gatchina, St.~Petersburg 188300,
Russia, \\
Th. Gutsche and Amand Faessler\\ Institut f\"ur Theoretische Physik,
Universit\"at T\"ubingen\\ T\"ubingen D-72076, Germany}
\date{}
\maketitle

\begin{abstract}

We calculate the nucleon parameters in nuclear matter using the QCD sum
rules approach in  gas approximation. Terms up to $1/q^2$ in
the operator product expansion (OPE) are taken into account. The higher
moments of the nucleon structure functions are included. The complete
set of the nucleon expectation values of the four-quark operators is
employed. Earlier the lack of information on these values has been the
main obstacle for the further development of the
approach.
We show that the values of the four-quark condensates are
consistent with the assumptions about the convergence of
the OPE. Inclusion of these condensates and of the
nonlocality of the vector condensate are important for the
calculation of the nucleon parameters.
The
nucleon vector self-energy $\Sigma_v$ and the nucleon effective mass
$m^*$ are expressed in terms of the in-medium values of QCD
condensates. The numerical results for these parameters at the
saturation value of the density agree with those obtained by the
methods of nuclear physics.
\end{abstract}

\section{Introduction}

The QCD sum rules were invented by Shifman {\em et al.} \cite{1} to
express the hadron parameters through the vacuum expectation values of
QCD operators. Being initially used for the mesons, the method was
expanded by Ioffe \cite{2} to the description of the baryons. The
approach succeeded in describing the static characteristics as well as
some of the dynamical characteristics of the hadrons in vacuum --- see,
\eg, the reviews \cite{3,4}.

The basic idea is to consider the
correlation function $\Pi_0(q^2)$ describing the
propagation of the system with the quantum numbers of the hadron, in
the different regions of values of the momentum $q^2$, where certain
information on its behavior is available. The asymptotic freedom of QCD
enables to present $\Pi_0(q^2)$ at $q^2\to-\infty$ as a power series
in $q^{-2}$ and
the QCD coupling $\alpha_s$. On the other hand, the imaginary
part of $\Pi_0(q^2)$ at $q^2>0$ can be described in terms of the
observable hadrons. This prompts to consider the dispersion relation
for the function $\Pi_0(q^2)$ \cite{1}
\begin{equation}
\Pi_0(q^2)\ =\ \frac1\pi\int\frac{\im\Pi_0(k^2)}{k^2-q^2}\ dk^2
\end{equation}
at $q^2\to-\infty$. The coefficients of the expansion of the left-hand
side (lhs) of the function $\Pi_0(q^2)$ in powers of $q^{-2}$ are the
expectation values of the local operators constructed of the quark and
gluon fields, which are called "condensates". Such presentation, known
as the operator product expansion (OPE) \cite{5} provides the
perturbative expansion of the short-distance effects, while the
nonperturbative physics is contained in the condensates. The usual
treatment of the right-hand side (rhs) of Eq. (1) consists in "pole +
continuum" presentation, in which the lowest lying pole is singled out
while the higher states are approximated by the continuum. Thus, Eq. (1)
ties the values of QCD condensates with the characteristics of the
lowest hadronic state. Such interpretation requires that the contribution
 of the pole to the rhs of Eq. (1) exceeds the contribution of the
continuum.

The OPE of the lhs of Eq. (1) becomes increasingly valid, when the
value of $|q^2|$ increases. On the other hand, the "pole + continuum"
model becomes more accurate when $|q^2|$ decreases. The important
assumption is that the two presentations are close in a certain
intermediate region of the values of $q^2$. To improve the overlap of the
QCD and the phenomenological descriptions, one usually applies certain
mathematical tools, \ie\ the Borel transform. The Borel transformed
dispersion relations (1) are known as QCD sum rules \cite{1,2}.

For example, the QCD sum rules for the nucleon provided a connection
between the nucleon mass and the scalar quark condensate $\langle0|\bar
qq|0 \rangle$ \cite{2}. Similar relations have been obtained for the
magnetic moments of the nucleons \cite{6}, etc.

Later the QCD sum rules were applied for
the investigation of modified nucleon
 parameters  in  nuclear matter \cite{7,8}. They
were based on the Borel-transformed dispersion relation for the function
$\Pi_m(q)$ describing the propagation of the system with the quantum
numbers of the nucleon (the proton) in  nuclear matter. Considering
nuclear matter as a system of $A$ nucleons with momenta $p_i$,
one introduces the vector
\begin{equation} p\ =\ \frac{\Sigma p_i}A\ ,
\end{equation}
which is thus $p\approx(m,0)$ in the rest frame of the matter. The
function $\Pi_m(q)$ can be presented as $\Pi_m(q)=\Pi_m(q^2,\varphi(p,q))$
with the arbitrary function $\varphi(p,q)$ being kept constant in the
dispersion relations in $q^2$.

The spectrum of the function $\Pi_m(q)$ is much more complicated than
that of the function $\Pi_0(q^2)$. The choice of the function
$\varphi(p,q)$ is dictated by  attempts to separate the singularities
connected with the nucleon in the matter from those connected with the
properties of the matter itself. Since the latter manifest themselves as
singularities in the variable $s=(p+q)^2$, the separation can be done by
putting $\varphi(p,q)=(p+q)^2$ and by fixing \cite{7,8,9}
\begin{equation}
\varphi(p,q)\ =\ (p+q)^2\ \equiv\ s\ =\ 4E_{0F}^2
\end{equation}
with $E_{0F}$ being the relativistic value of the nucleon energy at
the Fermi surface.

The general form of the function $\Pi_m$ can thus be presented as
\begin{equation}
\Pi_m(q)\ =\ q_\mu\gamma^\mu\Pi^q_m(q^2,s)+p_\mu\gamma^\mu\Pi^p_m
(q^2,s)+I\Pi^I_m(q^2,s)\ .
\end{equation}
The in-medium QCD sum rules are the Borel-transformed dispersion
relations for the components $\Pi^j_m(q^2,s)$ $(j=q,p,I)$
\begin{equation}
\Pi^j_m(q^2,s)\ =\ \frac1\pi\int \frac{\im\Pi^j_m(k^2,s)}{k^2-q^2}\
dk^2\ .
\end{equation}
It was shown in \cite{7,8,9,10}
that the "pole + continuum" model for the rhs of Eq. (5) can be used
at least until we do not include the higher order terms of the density
expansions of the functions $\Pi^j_m(q^2,s)$. Thus, one can expect
that the characteristics of the nucleon in nuclear matter can be
expressed through the in-medium values of QCD condensate.

In the lowest order of OPE the problem was approached in
\cite{7,8,9,10}.
It was noticed that the condensates of the lowest dimensions
$(d=3,4)$ can  either  be calculated or expressed through the
observables. The vector condensate $v_\mu(\rho)=\langle M|\sum_i
\bar q^i(0)\gamma_\mu q^i(0)|M\rangle$ is proportional to the density of
the matter $\rho$, being
\begin{equation}
v_\mu(\rho)=v_{N\mu}\rho\ , \qquad v_{N\mu}=\langle N|\sum_i
\bar q^i(0)\gamma_\mu q^i(0)|N\rangle\ .
\end{equation}
Here the upper index $i$ denotes the quark flavor.
In the rest frame of the matter
 we get $v_\mu(\rho)=v(\rho)\delta_{\mu0}$,
$v_{N\mu} =v_N\delta_{\mu0}$ with
$$
v(\rho)\ =\ v_N\rho\ ; \quad v_N\ =\ 3 \eqno{(6')}
$$
being just the number of the valence quarks in the nucleon. The scalar
condensate is
\begin{eqnarray}
&&\hspace*{-1cm} \kappa_m(\rho)\ =\ \langle M|\sum_i\bar
q^i(0)q^i(0)|M\rangle\ =\ \kappa_0 +\kappa(\rho)\ , \nonumber\\
&&\hspace*{-1cm}  \kappa_0=\kappa_m(0)\ , \quad \kappa(\rho)=\kappa_N
\rho+\cdots,
\quad \kappa_N=\langle N|\sum_i\bar q^i(0)q^i(0)|N\rangle\ .
\end{eqnarray}
Here the dots denote the terms which are nonlinear in
$\rho$. The expectation value $\langle N|\sum_i\bar q^iq^i|N\rangle$ is
related to the  $\pi N$ sigma term $\sigma_{\pi N}$, \ie\
\cite{11}
\begin{equation}
\kappa_N\ =\ \langle N|\bar uu+\bar dd|N\rangle\ =\
\frac{2\sigma_{\pi N}}{m_u+m_d}
\end{equation}
with $m_{u,d}$ standing for the current masses of the light quarks.

Turning to the condensates of dimension $d=4$, we find for the gluon
condensate \cite{7}
\begin{eqnarray}
&& g_m(\rho)\ =\ \langle M|\frac{\alpha_s}\pi\ G^2(0)|M\rangle\ =\
g_0+g(\rho)\ , \nonumber\\
&& g_0\ =\ g_m(0)\ , \quad g(\rho)\ =\ g_N\rho+\ \ldots
\end{eqnarray}
with the nucleon expectation value
\begin{equation}
g_N\ =\ \langle N|\frac{\alpha_s}\pi\ G^2(0)|N\rangle\ \approx\
-\frac89\,m\ ,
\end{equation}
obtained in \cite{12} in a model-independent way.

Also, the nonlocal condensate $\langle N|\bar q(0)\gamma_\mu q(x)|N
\rangle$ provides the contributions of $d=4$ and those of the higher
dimension. The term of the dimension $d=4$ is
\begin{equation}
\langle N|\bar q^i(0)\gamma_\mu D_\nu q^i(0)|N\rangle\ =\ \left(
g_{\mu\nu}-\frac{4p_\mu p_\nu}{m^2}\right)mx_2
\end{equation}
with $x_2$ standing for the second moment of the nucleon structure
function \cite{9}.
The nonlocality of the scalar operator $\bar q(0)q(x)$
manifests itself in the higher terms of the operator
 expansion. The nonlocality of the product of the gluon
operators is not expected to be important because of the
minor contribution of the gluon expectation value to the
nucleon parameters.

The shift of the position of the nucleon pole, which in the linear
approximation can be identified with the single-particle potential
energy of the nucleon, was expressed  as a linear combination of the
condensates of the lowest dimension \cite{7,8}. The vector and scalar
expectation values appeared to be the most important ingredients. Their
contributions cancelled to large extent, reproducing the familiar
features of the Walecka model \cite{13}. An alternative approach was
developed in \cite{14,15,16}
with the dispersion relations in the time component $q_0$ at
three-dimensional momentum $|{\bf q}|$ being fixed. It provided
a similar result.

The lack of knowledge about the in-medium expectation values of the
higher dimension became the obstacle for the development of both
approaches. One of such expectation values is the scalar four-quark
condensate $\langle M|\bar qq\bar qq|M\rangle$. It was noticed in
\cite{8,16} that the configuration $2\langle0|\bar qq|0\rangle\langle
N|\bar qq|N\rangle\rho$ (with $\rho$ standing for the
 baryon density)  is
one of those, which
composed the in-medium expectation value of
the operator $\bar qq\bar qq$. In the gas approximation the expectation
value of the colour-singlet operator is
\begin{equation}
\langle M|\bar qq\bar qq|M\rangle = \langle0|\bar qq\bar qq|0\rangle+
2\rho\ \langle0|\bar qq|0\rangle\langle N|\bar qq|N\rangle
+ \rho\ \langle N|(\bar qq\bar qq)_{int}|N\rangle
\end{equation}
with the last term describing the "internal" action of the operators
inside the nucleon. In the "ground-state saturation approximation"
(also called "factorization approximation") formulated in \cite{15} the
last term of the rhs of Eq. (12) vanishes. This would lead to the
change $\langle M|\bar qq\bar qq|M\rangle-\langle0|\bar qq\bar
qq|0\rangle=2\rho\langle0|\bar qq|0\rangle\langle N|\bar qq|N\rangle$
of the value of the scalar four-quark condensate. Assuming this
approximation one would be forced to conclude that the four-quark
scalar condensate plays the crucial role in QCD sum rules, causing
doubts on the convergence of OPE. The numerical results would
contradict the known nuclear phenomenology \cite{16}.

There have been some attempts to get rid of the contribution of the
four-quark condensates, applying the differential operators \cite{9} or
by choosing the form of the in-medium function $\Pi_m(q)$ which does
not couple to the four-quark condensates \cite{17}. However, some of
the information appeared to be lost in the former case, while there
still remained some unknown condensates in the latter case. Anyway,
there was no consistent analysis of  sum rules with the inclusion of
the four-quark condensates until now. On the other hand, there are some
indications that the second term of the rhs of Eq. (12) does not
provide the true scale for the in-medium modification of the value of
the scalar four-quark condensates. The calculations, carried out in
\cite{18} predicted strong cancellation between the second and the
third terms in the rhs of Eq. (12). Also, the arguments based on
chiral counting and supporting the violation of the in-medium
factorization have been given in \cite{19}.

In the present paper we build and solve the QCD sum rules in nuclear
matter in the gas approximation with the account of the condensates up
to the dimension $d=6$. This means that we include the terms of the
order $1/q^2$ of the OPE (recall that the leading OPE terms are of the
order $q^4\ln q^2$). This requires the inclusion of the four-quark
condensates $\bar u\Gamma^Xu\bar u\Gamma^Yu$, $d\Gamma^Xd\bar
d\Gamma^Yd$ and $\bar u\Gamma^Xu\bar d\Gamma^Yd$ with $\Gamma^{X,Y}$
standing for the basic $4\times4$ matrices, corresponding to the
scalar, pseudoscalar, vector, pseudovector (axial) and tensor
structures.

In the gas approximation the in-medium expectation
value of any operator
$\hat A$ is
\begin{equation}
\langle M|\hat A|M\rangle\ =\ \langle0|\hat A|0\rangle
 + \rho\langle N|\hat A|N\rangle
\end{equation}
with $|N\rangle$ standing for the
 state vector of the free
unpolarized nucleon. Since we include only  terms linear in $\rho$,
we can neglect the Fermi motion of the nucleons of the matter. Thus we
put
\begin{equation}
s\ =\ 4m^2
\end{equation}
in Eq. (3). Having in mind the future extension of the approach, we
shall keep the dependence on $s$, using Eq. (14) for the specific
computations.

We consider  symmetric nuclear matter with an equal density of
the protons and neutrons $\rho_p=\rho_n=\rho/2$.

The nucleon expectation values of the lowest dimensions can either
be calculated in a model-independent way or expressed through the
observables. The calculations of the four-quark condensates require
model assumptions on the structure of the nucleon. The complete set of
the four-quark condensates was obtained in \cite{20} by using
features of the
perturbative chiral quark model (PCQM). The chiral quark model,
originally suggested in \cite{21}, was developed further in \cite{22}.
In the PCQM the nucleon is treated as a system of relativistic valence
quarks moving in an effective static field. The valence quarks are
supplemented by a perturbative cloud of  pseudoscalar mesons, in
agreement with the requirements of the chiral symmetry. In \cite{20}
a simple version close to the SU(2) flavor PCQM,
 which includes only the pions, has been used.

There are three types of  contributions to the four-quark condensate
in the framework of this approach. All four operators can act on the
valence quarks. Also, four operators can act on the pion. There is also
a possibility that two of the operators act on the valence quarks while
the other two act on the pions. Following \cite{20} we speak of the
"interference terms" in the latter case.

To obtain the contribution of the pion cloud, we need the expectation
values of the four-quark operators in pions. The latter have been
deduced
in \cite{23} by using the current algebra technique. We obtain a
remarkable cancellation of the pion contributions in the function
$\Pi_m(q)$. This cancellation takes place in any model of the nucleon
which treats the pion cloud perturbatively. Thus, the contributions of
the four-quark condensates come from the terms, determined by the
valence quarks only and from the interference terms.

We find the contribution of the four-quark condensate to the in-medium
modification $\Pi_m-\Pi_0$ of the function $\Pi_0$ to be much smaller,
then one could expect by assuming the "in-medium factorization
approximation". These terms are about 4-5 times smaller than the
leading ones of the OPE series. This is consistent with the hypothesis
of the convergence of OPE.

Thus, we obtain three sum rules equations for the functions
$\Pi^q_m(q^2,s)$, $\Pi^p_m(q^2,s)$ and $\Pi^I_m(q^2,s)$
introduced in Eq.(4).  The in-medium characteristics of the
nucleon, \ie\ the vector self-energy $\Sigma_v$ and the
effective mass $m^*$ are the unknowns of these equations. There
are two unknown parameters more, \ie\ the residue at the nucleon
pole $\lambda^{*2}_m$ and the continuum threshold $W^2_m$. All
these characteristics will be obtained from the QCD sum rules.

The dependence of the rhs of Eq. (5) on the parameters, which are
expected to be determined, is not linear (except the dependence on
$\lambda^{*2}_m$). Thus, even in the gas approximation the behavior of
these parameters with $\rho$ is linear only at sufficiently small
values of the density. We consider  two approaches to the problem.
In the linearized case we determine only the linear parts of the
in-medium modifications of the hadron (nucleon) and continuum
parameters. We construct the combination of the sum rules for the
function $\Pi_m-\Pi_0$ in such a way, that  two of the
equations determine
the values of $\Sigma_v$ and $m^*-m$ separately. Note that it is
possible to write the equation in which the parameter $m^*-m$ is the
only unknown, only because the proton has a definite space parity.
The third equation enables to find the in-medium changes of the
parameters $\delta\lambda^2=\lambda^{*2}_m -\lambda^2_0$ and $\delta
W^2=W^2_m-W^2_0$. We express the nucleon characteristics $\Sigma_v$ and
$m^*-m$ in terms of the vector, scalar, gluon and four-quark
condensates and of the moments of the structure functions.

In the nonlinearized version we do not assume the in-medium changes of
the parameters to be small. The three equations for the Borel
transformed function $\Pi_m(q)-\Pi_0(q^2)$ enable to obtain the values
of $\Sigma_v$, $\lambda^{*2}_m$ and $W^2_m$. At  densities $\rho$
of the order of the  saturation value $\rho_0$ the values of
$\Sigma_v$ and $m^*-m$ appear to coincide within 25$\%$ and
$10\%$ accuracy with the values provided by the linear version.
This causes somewhat larger difference in the values of the
potential energy $U(\rho)$ which still has reasonable values.

Inclusion of the  four-quark condensates and of
the higher moments of the structure functions
diminish the OPE value of the nucleon vector self-energy
 $\Sigma_v$ by about $25\%$ each. As to the scalar self-energy
$m^*-m$, the four-quark condensates
provide contribution of the same order as the leading
OPE term. However, this contribution is almost totally
compensated by the account of the higher moments of
the structure functions. Thus, the value of the $m^*-m$ is
very close to that given by the leading order of OPE.

The structure of the paper is as follows. In Sec. II we present
the sum rules in a form, which is convenient for our analysis.
In Sec. III and IV we calculate the contribution of the
dimension $d=6$, \ie\  the expansion of the nucleon structure
functions and the four-quark condensates. In Sec. V - VII we
present the solutions in the linearized and nonlinearized forms.
We discuss and summarize the results in Sec. VIII and IX.

\section{General equations}
\subsection{Sum rules in vacuum}

To make the paper self-consistent, we recall the main points of the QCD
sum rules
 approach in vacuum \cite{1,2}. The function $\Pi_0(q^2)$ (often
referred to as "polarization operator") is presented as
\begin{equation}
\Pi_0(q^2)\ =\ i\int d^4xe^{i(qx)}\langle0|Tj(x)\bar j(0)|0\rangle
\end{equation}
with $j$ being the three-quark local operator
 (often referred to as "current")
with the proton quantum
numbers. The usual choice is \cite{2}
\begin{equation}
j(x)\ =\ \varepsilon^{abc}\left[u^{aT}(x)C\gamma_\mu u^b(x)\right]
\gamma_5\gamma_\mu d^c(x)\ ,
\end{equation}
where $T$ denotes a transpose and $C$ is the charge  conjugation
matrix. The upper indices denote the colors.

The lhs of Eq. (1) is approximated by several lowest terms of OPE, \ie\
$\Pi_0(q^2)\approx\Pi^{OPE}_0(q^2)$. The empirical data are used for
the spectral function $\im\Pi_0(q^2)$ on the rhs of Eq. (1). Namely, it
is known, that the lowest lying state is the bound state of three
quarks, which manifests itself as a pole in the (unknown) point
$k^2=m^2$. Since the next singularity is the branching point
$k^2=W^2_{ph}=(m+m_\pi)^2$, one can present
\begin{equation}
\im\Pi_0(k^2)\ =\ \lambda^2_N\delta(k^2-m^2)+f(k^2)\theta(k^2-W^2_{ph})
\end{equation}
with $\lambda^2_N$ being the residue at the nucleon pole. Thus, Eq. (1)
takes the form
\begin{equation}
\Pi^{OPE}_0(q^2)\ =\ \frac{\lambda^2_N}{m^2-q^2}+\frac1\pi
\int\limits^\infty_{W^2_{ph}} \frac{f(k^2)}{k^2-q^2}\ dk^2\ .
\end{equation}
Of course, the detailed structure of the spectral density $f(k^2)$
cannot be resolved in such an approach. The further approximations are
based on the asymptotic behavior
\begin{equation}
f(k^2)\ =\ \frac1{2i}\ \Delta\Pi^{OPE}_0(k^2)
\end{equation}
at $k^2\gg|q^2|$ with $\Delta$ denoting the discontinuity. The
discontinuity is caused by the logarithmic contributions of the
perturbative OPE terms. The usual ansatz consists in extrapolation of
Eq. (19) to all the values of $k^2$, replacing also the physical
threshold $W^2_{ph}$ by the unknown effective threshold $W^2_0$, \ie
\begin{equation}
\frac1\pi\int\limits^\infty_{W^2_{ph}}\frac{f(k^2)}{k^2-q^2}\ dk^2\ =\
\frac1{2\pi i} \int\limits^\infty_{W^2_0}
\frac{\Delta\Pi^{OPE}_0(k^2)}{k^2-q^2}\ dk^2\ .
\end{equation}
Thus Eq. (1) takes the form
\begin{equation}
\Pi^{OPE}_0(q^2)\ =\ \frac{\lambda^2_N}{m^2-q^2} +
\frac1{2\pi i}\int\limits^\infty_{W^2_0}\frac{\Delta\Pi^{OPE}_0
(k^2)}{k^2-q^2}\ dk^2\ .
\end{equation}

The lhs of Eq. (21) contains QCD condensates. The rhs of Eq. (21)
contains three unknown parameters: $m,\lambda^2_N$ and $W^2_0$. The OPE
becomes increasingly true when the value $|q^2|$ increases. The "pole +
continuum" model is more accurate at the smaller values of $|q^2|$.
Thus one can expect Eq. (21) to be true in a certain limited interval
of the values of $|q^2|$. To improve the overlap of the OPE and the
phenomenological description one usually applies the Borel transform
defined as
\begin{eqnarray}
&& Bf(q^2)\ =\ \lim\limits_{Q^2,n\to\infty}\frac{(Q^2)^{n+1}}{n\,!}
\left(-\frac d{dQ^2}\right)^nf(q^2)\ \equiv\ \tilde f(M^2)\ ,
\nonumber\\
&& Q^2\ =\ -q^2; \quad M^2\ =\ Q^2/n
\end{eqnarray}
with $M$ called  Borel mass. It is important in the applications to
the sum rules that the Borel transform eliminates the polynomials and
emphasizes the contribution of the lowest state in rhs of Eq. (21) due
to the relation
\begin{equation}
B\ \frac1{m^2-q^2}\ =\ e^{-m^2/M^2}\ .
\end{equation}
The Borel-transformed form of Eq. (21) reads
\begin{equation}
\widetilde\Pi^{OPE}_0(M^2)\ =\ \lambda^2_Ne^{-m^2/M^2}+\frac1{2\pi i}
\int\limits^\infty_{W^2_0}dk^2e^{-k^2/M^2}\Delta\,\Pi^{OPE}_0 (k^2)
\end{equation}
and is known as QCD sum rules. Actually, there are two sum rules for
the structures $\Pi^q_0$ and $\Pi^I_0$ of the function
$\Pi_0(q)=q_\mu\gamma^\mu\Pi^q_0(q^2)+I\Pi^I_0(q^2)$ with $I$ standing
for the unit matrix.

It appeared to be more convenient to work with Eq. (24) multiplied by
the numerical factor $32\pi^4$. The two sum rules for the nucleon in
vacuum can be presented in the form \cite{2}
\begin{eqnarray}
&& L^q_0(M^2,W^2_0)\ =\ \Lambda_0(M^2)\ . \\
&& L^I_0(M^2,W^2_0)\ =\ m\Lambda_0(M^2)
\end{eqnarray}
with
\begin{equation}
\Lambda_0(M^2)\ =\ \lambda^2_0 e^{-m^2/M^2}\ .
\end{equation}
Here $\lambda^2_0=32\pi^4\lambda^2_N$,
\begin{eqnarray}
&& L^q_0(M^2,W^2_0)=32\pi^4\bigg(\widetilde\Pi_0\,^{q,OPE}(M^2) -
\frac1{2\pi i}
\int\limits^\infty_{W^2_0}dk^2e^{-k^2/M^2}\Delta\,
\Pi^{q,OPE}_0(k^2)\bigg), \nonumber\\
&& L^I_0(M^2,W^2_0)\ =\
32\pi^4\bigg(\widetilde\Pi_0\,^{I,OPE}(M^2) - \frac1{2\pi i}
\int\limits^\infty_{W^2_0}dk^2 e^{-k^2/M^2}\Delta\,
\Pi^{I,OPE}_0(k^2)\bigg) .
\end{eqnarray}

The lhs of Eqs. (25) and (26) \cite{2,6} have been obtained by
including
 of the condensates of dimension $d=8$, \ie\ with the account of
the terms of the order $1/q^4$ in OPE of the functions $\Pi^{q,OPE}_0$
\begin{eqnarray}
L^q_0(M^2,W^2_0) &=& \frac{M^6E_2}{L^{4/9}}+
\frac{bE_0 M^2}{4L^{4/9}}+\frac43 a^2L^{4/9}
- \frac13 \frac{\mu^2_0}{M^2}a^2,\\
L^I_0(M^2,W^2_0) &=& 2aM^4E_1
-\frac{ab}{12}+\frac{272}{81}\,
\frac{\alpha_s}\pi\,a^3 \frac1{M^2}
\end{eqnarray}
with the
traditional notations $a=-(2\pi)^2\langle0|\bar
qq|0\rangle=-2\pi^2\kappa_0$ (we assumed the isotopic invariance
$\langle0|\bar uu|0\rangle=\langle0|\bar
dd|0\rangle=\langle0|\bar qq|0\rangle$), $b=(2\pi)^2g_0$,
$\mu^2_0=0.8\,\rm GeV^2$. Here $E_i$ are the functions of the ratio
$W_0^2/M^2$: $E_i=E_i(W_0^2/M^2)$. They are given by the formulas
\begin{equation}
E_0(x)=1-e^{-x}, \quad E_1(x) = 1-(1+x)e^{-x},\quad
 E_2(x) = 1-\left(1+x +\frac{x^2}2\right)e^{-x} .
\end{equation}
The factor
\begin{equation}
L(M^2) \ =\ \frac{\ln M^2/\Lambda^2}{
\ln\nu^2/\Lambda^2}
\end{equation}
accounts for the anomalous dimension, \ie\
the most important corrections of the order $\alpha_s$ enhanced by the
"large logarithms". In Eq. (32) $\Lambda=\Lambda_{QCD}=0.15\,$GeV,
while $\nu=0.5\,$GeV is the normalization point of the characteristic
involved. Note that the two last terms on the rhs of Eq. (29) originate
from the four-quark condensates $\langle0|\bar u\Gamma^X u\bar
u\Gamma^X u|0\rangle$ and can be expressed through the single term
$(\langle0|\bar qq|0\rangle)^2$ only in framework of the factorization
hypothesis \cite{1,2}. Also, the last term on the rhs of Eq. (30) is
the six-quark condensate, evaluated in the same approximation.

The matching of the lhs and rhs of Eqs. (25) and (26) have been
achieved \cite{2,6} in the domain
\begin{equation}
0.8\mbox{ GeV}^2\ <\ M^2\ <\ 1.4\,\rm GeV^2
\end{equation}
providing the values of the vacuum parameters
$$
\lambda^2_0=1.9\mbox{ GeV}^6\ ; \quad W^2_0=2.2\mbox{ GeV}^2
\eqno{(33')}
$$
if $m=0.94$ GeV.

\subsection{Sum rules in nuclear matter}

The OPE terms of the polarization operator in nuclear matter
\begin{equation}
\Pi_m(q)\ =\ i\int d^4 xe^{i(qx)}\ \langle M|Tj(x)\bar j(0)|M\rangle
\end{equation}
contains the in-medium values of QCD condensates. Some of these
condensates vanish in the
vacuum, obtaining non-zero values only in the medium.
The other ones just change their values compared to the vacuum ones.

The spectrum of the function $\Pi_m(q)$ is much more complicated, than
that of the vacuum function $\Pi_0(q^2)$. However,
\cite{7,8,9,10}
the spectrum of the function $\Pi_m(q^2,s)$ at fixed value of $s$ can
be described by the "pole + continuum" model at least until we include
the terms of the order $\rho^2$ in the OPE of $\Pi_m(q^2,s)$.

The description of the nucleon pole is based on the general expression
for the propagator
\begin{equation}
G^{-1}_N\ =\ \left(G^0_N\right)^{-1} - \Sigma
\end{equation}
with $G^0_N=(q_\mu\gamma^\mu-m)^{-1}$ being the propagator of the free
nucleon, while
\begin{equation}
\Sigma\ =\ q_\mu\gamma^\mu\Sigma_q+\frac1m\ p_\mu\gamma^\mu\Sigma_p
+\Sigma_s
\end{equation}
is the general form of the self-energy of the nucleon in nuclear
mater. In the kinematics, determined by Eq. (3) we obtain
\begin{equation}
G_N\ =\ Z\cdot\frac{q_\mu\gamma^\mu-p_\mu\gamma^\mu
(\Sigma_v/m)+m^*}{q^2-m^2_m}
\end{equation}
with
\begin{equation}
\Sigma_v\ =\ \frac{\Sigma_p}{1-\Sigma_q}\ , \quad
m^*\ =\ \frac{m+\Sigma_s}{1-\Sigma_q}\ .
\end{equation}
The new position of the nucleon pole is
\begin{equation}
m^2_m\ =\ \frac{(s-m^2)\Sigma_v/m-\,\Sigma^2_v+m^{*2}}{
1+\Sigma_v/m}\ ,
\end{equation}
while
\begin{equation}
Z\ =\ \frac1{(1-\Sigma_q)(1+\Sigma_v/m)}\ .
\end{equation}

Thus, we shall present the dispersion relations for the functions
$\Pi^i_m(q^2,s)$ $(i=q,p,I)$ determined by Eq. (4) in the form
\begin{equation}
\Pi^{i,OPE}_m(q^2,s)=\frac{Z\lambda^2_m b_i}{m^2_m-q^2}+\frac1{2\pi i}
\int\limits^\infty_{W^2_m} \frac{\Delta_{k^2}\Pi^{i,OPE}(k^2,s)}{
k^2-q^2}
\end{equation}
with $b_q=1$, $b_p=-\Sigma_v$, $b_I=m^*$. The Borel-transformed
sum rules take the form
\begin{eqnarray}
\label{45}
L^q_m(M^2,W^2_m) &=& \Lambda_m(M^2)\ ;\\
\label{46}
L^p_m(M^2,W^2_m) &=& -\Sigma_v\Lambda_m(M^2)\ ;\\
\label{47}
L^I_m(M^2,W^2_m) &=& m^*\Lambda_m(M^2)
\end{eqnarray}
with
\begin{equation} \label{48}
\Lambda_m(M^2)\ =\ \lambda^{*2}_m e^{-m^2_m/M^2}\ .
\end{equation}
Here
\begin{equation} \label{49}
\lambda^{*2}_m\ =\ \lambda^2_m\cdot Z
\end{equation}
is the effective value of the residue in nuclear matter.

We present the lhs of Eqs. (\ref{45})--(\ref{47}) as
\begin{equation} \label{50}
L^i_m\ =\ \ell^i_m+u^i_m+\omega^i_m
\end{equation}
with $\ell^i_m(M^2,W^2_m)$ standing for the lowest order OPE terms,
$u^i_m(M^2,W^2_m)$ denoting the contribution of the higher moments of
the structure functions, while $\omega^i_m(M^2)$ provides the
contribution of the four-quark condensates. We write, correspondingly,
$L^i_0=\ell^i_0+\omega^i_0$ for the lhs of the vacuum sum rules
presented by Eqs. (25) and (26). We present also
\begin{eqnarray}
&& \ell^i(M^2,W^2_m)\ =\ \ell^i_m(M^2,W^2_m)-\ell^i_0(M^2,W^2_m)\ ,
\nonumber\\
&& \omega^i(M^2)\ =\ \omega^i_m(M^2)-\omega^i_0(M^2)\ .
\label{51}
\end{eqnarray}

In these notations the lowest order OPE terms are
\begin{equation}
\ell^q_0=\frac{M^6E_{2m}}{L^{4/9}}+\frac14\,\frac{bM^2E_{0m}}{L^{4/9}},
\quad \ell^p_0=0, \quad \ell^I_0=2aM^4E_{1m}
\label{52}
\end{equation}
and
\begin{eqnarray}
\ell^q &=& f^q_v(M^2,W^2_m)v(\rho)+f^q_g(M^2,W^2_m)g(\rho)\ ,
\nonumber\\
\label{53}
\ell^p &=& f^p_v(M^2,W^2_m)v(\rho),
\nonumber\\
\ell^I &=& f^I_\kappa(M^2,W^2_m)\kappa(\rho)
\nonumber\\
\end{eqnarray}
with
\begin{eqnarray}
f^q_v &=& -\frac{8\pi^2}3\, \frac{(s-m^2)M^2E_{0m}-M^4E_{1m}}{mL^{4/9}},
\quad f^q_g=\frac{\pi^2 M^2E_{0m}}{L^{4/9}}\ ,
\nonumber\\
f^p_v &=& -\frac{8\pi^2}3\,\frac{4M^4E_{1m}}{L^{4/9}}\ ,
\nonumber\\
f_\kappa^I &=& -4\pi^2M^4E_{1m}\ .
\label{54}
\end{eqnarray}
The functions $v(\rho)$, $\kappa(p)$, $g(\rho)$ are determined by Eqs.
(6'), (7), (9). The notation $E_{km}$ $(k=0,1,2)$ means that the
functions depend on the ratio $W^2_m/M^2$. Actually, the higher moments
of the structure functions of the nucleon have been neglected in Eqs.
(\ref{53}) and (\ref{54}).

\section{
Accounting for x-dependence of the operators.\newline
Contributions of the higher moments and \newline of the higher
twists  of the structure functions}

The calculation of the function $\Pi_m(q^2,s)$ defined by Eq. (34) is
based on the presentation of the single-quark propagator in the medium
\begin{equation}
 \langle M|Tq^i_\alpha(x)\bar q\,^i_\beta(0)|M\rangle =
G_{\alpha\beta}(x)-\frac14 \langle M|\bar q\,^i(0)\gamma_\mu
q^i(x)|M\rangle\gamma^\mu_{\alpha\beta}-\frac14\langle M|
\bar q\,^i(0)q^i(x)|M\rangle \delta_{\alpha\beta}
\label{55}
\end{equation}
with $G(x)=(ix_\mu\gamma_\mu)/(2\pi^2x^4)$ being the free propagator
of the quark in the chiral limit.
Recall that $i$ denotes the light quark flavor.
 In the lowest orders of OPE two of the quarks are described by
the free propagators and only one of the quarks is presented by the
second or the third term of the rhs of Eq. (\ref{55}).

At $x=0$ the matrix elements in the second and third terms on the rhs
are just the vector and scalar condensates defined by Eqs. (6) and (7).
The  contribution of the bilocal configurations can be expressed
in terms of the higher moments and twists of the nucleon structure
functions \cite{9}.

The bilocal operators on the rhs of Eq. (\ref{55}) are not gauge
invariant.  The gauge  invariant expression, achieved by substitution
\cite{24}
\begin{equation}
q^i(x)\ =\ q^i(0)+x_\alpha D_\alpha q^i(0)+\frac12\,x_\alpha x_\beta
D_\alpha D_\beta q^i(0)+ \cdots
\label{56}
\end{equation}
with $D_\alpha$ standing for the covariant derivatives, provides the
infinite set of the local condensates. The expectation values depend on
the variables $(px)$ and $x^2$. In the gas approximation we  only need
the nucleon matrix elements Eq. (13). For the vector
structure the general form is \begin{equation} \label{57}
\theta^i_\mu(x)=\langle N|\bar q^i(0)\gamma_\mu q^i(x)|N\rangle =
\frac{p_\mu}m\,\phi^i_a ((px),x^2)+ix_\mu m\phi^i_b((px),x^2)
\end{equation}
with $q^i(x)$ defined by Eq. (\ref{56}).

Expansion in powers of $x^2$ corresponds to the expansion of the
function $\Pi_m(q)$ in powers of $q^2$. To obtain the terms of the
order $q^{-2}$ it is sufficient to include two lowest terms of the
expansions in powers of $x^2$. One can present \cite{9,25}
\begin{equation}
\label{58}
\phi^i_{a(b)}((px),x^2)\ =\ \int\limits^1_0 d\alpha
e^{-i\alpha(px)}f^i_{a(b)}(\alpha,x^2)
\end{equation}
with
\begin{equation}\label{58a}
f^i_{a(b)}(\alpha,x^2)=\eta^i_{a(b)}(\alpha) + \frac18
x^2m^2\xi^i_{a(b)}(\alpha) .
\end{equation}
Here $\eta^i_a(\alpha)=f^i_a(\alpha,0)$ is
the contribution of the quarks with the flavor $i$ to the asymptotics
of the nucleon structure function $\eta(\alpha)=\eta^u_a(\alpha)
+\eta^d_a(\alpha)$, normalized by the condition
\begin{equation} \label{59}
\int^1_0 d\alpha\eta(\alpha)\ = \ 3
\end{equation}
with the rhs presenting just the number of the valence quarks in the
nucleon. Thus, expansion of the function
$\varphi^i_a(px)=\phi^i_a((px),0)$ in powers of $(px)$ is expressed
through the moments of the distributions $\eta^i_a(\alpha)$. The
moments are well known --- at least, those, which are numerically
important. Also, the first moment of the distribution
$\xi_a(\alpha)=\xi^u_a(\alpha)+\xi^d_a(\alpha)$
\begin{equation}
\label{60}
\xi\ =\
\int\limits^1_0\left(\xi^u_a(\alpha)+\xi^d_a(\alpha)\right)d\alpha\
\approx\ -0.3
\end{equation}
was calculated in \cite{26} by QCD sum rules method. The moments
of the function $\eta^i_b(\alpha)$ can be obtained by using the
equations of motion $D_\alpha\gamma^\alpha q^i(x)=m_iq^i(x)$. Thus, in
the chiral limit \cite{9}
\begin{eqnarray}
&& \langle\varphi^i_b\rangle =\frac14\langle\varphi^i_a\alpha\rangle\ ,
\nonumber\\
&& \langle\varphi^i_b\alpha\rangle=\frac15\left(\langle\varphi^i_a
\alpha^2\rangle -\frac14\langle\xi^i\rangle \right),
\nonumber\\
&&  \langle \xi^i_b\rangle\ =\ \frac16\langle \xi^i_a\alpha\rangle\ .
\label{61}
\end{eqnarray}
Here we denoted
\begin{equation}\label{e59}
\langle f\rangle=\int^1_0 d\alpha f(\alpha)
\end{equation}
for any function $f(\alpha)$.

Note that the nonlocality of the scalar condensate, \ie\ of the last
term on the rhs of Eq. (\ref{55}) does not manifest itself in the terms
up to $1/q^2$. The first derivative in $(px)$, as well as all the
derivatives of the odd order vanish in the chiral limit due to QCD
equation of motion. The next to leading order of the expansion in
powers of $x^2$ vanishes due to certain cancellations \cite{8} as well
as in the case of vacuum \cite{2} for the particular choice of the
operator $j(x)$ presented by Eq. (16).
We do not account for the nonlocality of the gluon
operators, since the gluon  expectation values
play the minor role in our sum rules.

Now we are ready to calculate the contributions
$\Pi_{nl}(q)$ of the nonlocal vector condensate to the polarization
operator $\Pi_m(q)$. We express $\Pi_{nl}$ in terms of the proton
expectation values
$\theta^i_\mu(x)=\langle p|\bar q^i(0)\gamma_\mu q^i(x)|p\rangle$.
Employing the isotopic invariance we obtain
\begin{equation} \label{62}
\Pi_{nl}(q)=
\frac{4i}{\pi^4}\int\frac{d^4x}{x^8}\left(x^2
\frac{\hat\theta^u + \hat\theta^d}{2}
+ \hat x(x,\theta^u+\theta^d)\right)e^{i(qx)}\cdot\rho,
\end{equation}
contributing to the vector structures $\hat q$ and $\hat p$ of the
polarization
operator $\Pi_m(q)$. Here we denoted
\begin{equation}\label{e61}
\hat a = a_\mu\gamma^\mu .
\end{equation}
Using Eq.(\ref{57}) we obtain
\begin{equation}\label{e62}
\Pi_{nl}(q) = \Pi^a_{nl}(q) + \Pi^b_{nl}(q)
\end{equation}
with
\begin{equation}\label{e63}
\Pi_{nl}^a(q)=\frac{4i}{\pi^4}
\int\frac{d^4x}{x^8}\left(x^2\hat p
\frac{\phi^u_a + \phi^d_a}{2}
+ \hat x(xp)(\phi^u_a+\phi_a^d)\right)e^{i(qx)}\cdot\rho
\end{equation}
$$
\Pi_{nl}^b(q)= - \frac{6m}{\pi^4}
\int\frac{d^4x}{x^6}\hat x
(\phi^u_b+\phi_b^d)e^{i(qx)}\cdot\rho
$$
We present each of the terms $\Pi_{nl}^{a(b)}$ as the sum
$\Pi_{nl}^{1a(b)} + \Pi_{nl}^{2a(b)}$, corresponding to the two
terms of the expansion in powers of $x^2$ in Eq.(\ref{58a}).
In particular, the contribution $\Pi_{nl}^{1a}$, which is  numerically
most important, can be presented as:
\begin{equation}
\label{63}
\Pi^{1a}_{nl}(q)=\left[\frac1{6m\pi^2}\int\limits^1_0 d\alpha\hat q\,'
(pq')\ln(\frac{-q^{'2}}{\Lambda_c^2})\eta_a(\alpha)
+ \frac{\hat p}{3m\pi^2}\int\limits^1_0 d\alpha q^{'2}
\ln(\frac{-q^{'2}}{\Lambda_c^2})\eta_a(\alpha)\right]\rho
\end{equation}
with $\eta_a(\alpha)=\eta^u_a(\alpha)+\eta^d_a(\alpha)$,
$q'=q-p\alpha$ (see Appendix A). From Eqs. (3) and (14) one finds
$(pq)=(s-m^2-q^2)/2$. The cutoff $\Lambda_c$ will be eliminated by the
Borel transform.

Presenting $q'^2 = -(1+\alpha)(Q^2+A^2)$ where $Q^2=-q^2$,
$A^2(\alpha) = \alpha (s-m^2-m^2\alpha)/(1+\alpha)$
we see that the second term of the expression
$\ln q'^2=\ln q^2 + \ln(\frac{q'^2}{q^2})$ does not have a cut,
running to infinity, but has a finite cut. This singularity
requires a special treatment in QCD sum rules. On the other hand,
it is the singularity in the $u$ channel of the interaction of the
baryon current with the quark of the nucleon of matter. It
corresponds to the exchange terms on the rhs  of the sum rules. In
this paper we neglect the nonlocal singularities, thus claiming for the
description of the nucleon in the Hartree approximation. However,
we include the regular smooth dependence on the higher moments.

The contributions $\Pi_{nl}^{1b}$
and $\Pi_{nl}^{2a(b)}$ can be expressed
in terms of the moments of the functions $\eta^i_b$ and
$\xi^i_{a,b}$ (see Appendix A). Since the higher moments of the
functions $\eta^i_a(\alpha)$, as well as the value of $\xi$ are
small, we include only the lowest moments of the functions
$\eta^i_b(\alpha)$ and the first moments of the functions
$\xi^i_a(\alpha)$ --- Eq. (\ref{61}). The last of the equalities
(\ref{61}) enables us to neglect the contribution of the functions
$\xi^i_b(\alpha)$.

Finally, the higher moments and higher twists of the nucleon
structure functions provide the contributions $u^i$ to the lhs
$L^i_m$ of the sum rules --- Eq. (\ref{50})
\begin{eqnarray}
&& u^q(M^2)\ =\ u^q_N(M^2)\rho\ ; \nn \\
&& u^q_N(M^2)=\frac{8\pi^2}{3L^{4/9}m}\left[
-\frac52 m^2 M^2 E_{0m} \langle\eta\alpha\rangle
+\frac32 m^2 (s-m^2)\langle\xi\rangle\right] ;
\nn \\
&&  u^p(M^2)\ =\ u^p_N(M^2)\rho\ ;
\label{68}\\
&& u^p_N(M^2)=\frac{8\pi^2}{3L^{4/9}}\left[
-5(M^4 E_{1m}-(s-m^2)M^2 E_{0m})\langle\eta\alpha\rangle
\right.
\nn\\
&&\
\left.
-\frac{12}5 m^2M^2E_{0m}\langle\eta\alpha^2\rangle
+\frac{18}5 m^2M^2E_{0m}\langle\xi\rangle
\right] ;
\nn\\
&&  u^I(M^2)\ =\ 0\ .
\nn
\end{eqnarray}
Here we denote $L=L(M^2)$.
 Parameter $\xi$ is defined
by Eq. (\ref{60}).

\section{Contribution of the four-quark condensates}

The four-quark expectation values contribute to the OPE terms $1/q^2$
 of the function $\Pi_m(q)$. Now only one quark is determined by the
free propagator $G_q(x)$. Two other quarks are described by the last
term of the two-quark propagator
\begin{eqnarray}
&&\langle M|Tq_\alpha(x)\bar q_\beta(0)q_\rho(x)\bar q_\tau(0)|M\rangle
=[G_q(x)]^2-\frac14\langle M|\bar q\Gamma^Xq|M\rangle G_q(x)
\Gamma^X_{\alpha\beta}
\nn \\
- &&\ \frac14\langle M|\bar q\Gamma^Xq|M\rangle G_q(x)
\Gamma^X_{\rho\tau}+\frac1{16}\langle M|\bar q\Gamma^Xq\bar q\Gamma^Y
q|M\rangle\Gamma^X_{\alpha\beta}\Gamma^Y_{\rho\tau}
\label{69}
\end{eqnarray}
with $\Gamma^{X,Y}$ being the basic $4\times4$ matrices
\begin{equation} \label{70}
\Gamma^I=I, \quad \Gamma^{Ps}=\gamma_5, \quad \Gamma^V=\gamma_\mu, \quad
\Gamma^A=\gamma_\mu\gamma_5, \quad \Gamma^T=\frac i2(\gamma_\mu
\gamma_\nu-\gamma_\nu\gamma_\mu),
\end{equation}
acting on the Lorentz indices of the quark operators. Equation
(\ref{69}) is analogous to Eq. (\ref{55}) for the single-quark
propagator. We did not display the color indices in Eq. (\ref{69}),
keeping in mind that the quark operators are color antisymmetric
--- Eq. (16). One can write an equation similar to Eq.
(\ref{69}) for the quarks of different flavors.

Introducing the notations
\begin{equation} \label{71}
H^{XY}_m(\rho) = \langle M|\bar u\Gamma^Xu\bar u\Gamma^Yu|M
\rangle; \quad R^{XY}_m(\rho)=\langle M|\bar d\Gamma^Xd\bar u
\Gamma^Yu|M\rangle
\end{equation}
we write in the gas approximation
\begin{equation} \label{72}
H^{XY}_m(\rho)=H^{XY}_m(0)+\rho h^{XY}; \quad R^{XY}_m(\rho)=
R^{XY}_m(0)+\rho r^{XY}\ .
\end{equation}
The characteristics $h^{XY}$ and $r^{XY}$ can be presented as
\begin{eqnarray}
h^{XY}&=&\frac56\left(\langle0|\bar u\Gamma^Xu|0\rangle\langle N|
\bar u\Gamma^Yu|N\rangle+\langle0|\bar u\Gamma^Yu|0\rangle\langle
N|\bar u\Gamma^Xu|N\rangle\right)
\nn\\
\label{73}
&+& \langle N|(\bar u\Gamma^Xu\cdot\bar u\Gamma^Yu)_{int}|N\rangle\ ;
\\
r^{XY} &=&\frac23\left(\langle0|\bar d\Gamma^Xd|0\rangle\langle
N|\bar u\Gamma^Yu|N\rangle+\langle0|\bar u\Gamma^Yu|0\rangle\langle
 N|\bar d\Gamma^Xd|N\rangle\right)
\nn \\
&+& \langle N|(\bar d\Gamma^Xd\cdot \bar u\Gamma^Yu)_{int}|N\rangle\ .
\label{74}
\end{eqnarray}
Here the lower index $"int"$ means that all the four operators
are acting inside the nucleon. The coefficients 5/6 and 2/3 on rhs of
Eqs. (\ref{73}) and (\ref{74}) present the weights of the
color-antisymmetric states --- see Appendix B. These equations
are consistent with Eq. (13) if we assume that some of the
single-particle operators which compose the operator $\hat A$
can act on the vacuum state vector  --- see also \cite{10}.

The contribution of the four-quark expectation values to the in-medium
change of the polarization operator can be written as
\begin{equation}
\label{75}
(\Pi)_{4q}\ =\ (\Pi_m)_{4q}-(\Pi_0)_{4q}=\ \frac\rho{q^2}\left(
\sum_{X,Y}\mu_{XY}h^{XY}+\sum_{X,Y}\tau_{XY}r^{XY}\right).
\end{equation}
Here $\mu_{XY}$ and $\tau_{XY}$ are certain matrices in Dirac space.
They can be obtained by using the general expression for the function
$\Pi_m(q)$ presented in \cite{16}
\begin{eqnarray}
\mu_{XY} &=& \frac{\theta_Y}{16}\mbox{ Tr }(\gamma_\alpha
\Gamma^X\gamma_\beta\Gamma^Y)\gamma_5\gamma^\alpha\hat
q\gamma^\beta\gamma_5\ ;
\nn \\
\tau_{XY} &=& \frac{\theta_Y}4\mbox{ Tr }(\gamma_\alpha\hat q
\gamma_\beta\Gamma^Y)\gamma_5\gamma^\alpha\Gamma^X\gamma^\beta\gamma_5\
,\quad \hat q=q_\mu\gamma^\mu\ .
\label{76}
\end{eqnarray}
Here $\theta_Y=1$ if $\Gamma^Y$ has a vector or tensor structure,
while $\theta_Y=-1$ in the scalar, pseudoscalar and axial cases. The
sign is determined  by that of the commutator between matrix
$\Gamma^Y$ and the charge conjugation matrix $C$ --- Eq. (16).

The products $\mu_{XY}h^{XY}$ obtain nonzero values if the matrices
$\Gamma^X$ and $\Gamma^Y$ have the same Lorentz structure. In this case
all the structures presented by Eq. (\ref{70}) contribute to
$(\Pi_m)_{4q}$. The products $\tau_{XY}r^{XY}$  do not turn to zero
only if $\Gamma^Y$ has a vector or axial structure. In the latter case
$\Gamma^X$ should be an axial matrix as well. In the former case
$\Gamma^X$ can be either Lorentz scalar or Lorentz vector.

We denote $h^{XX}=h^X$, $\mu_{XX}=\mu_X$ and $r^{XX}=r^X$,
$\tau_{XX}=\tau_X$ for the similar Lorentz structures $X$ and $Y$. The
scalar and pseudoscalar expectation values are Lorentz scalars. Thus,
their contributions can be expressed through single parameters. The
latter is true also for the scalar-vector expectation value $r^{SV}$.
We obtain
\begin{equation}\label{77}
\mu_S=-\frac {\hat q}{2}\ ,
\quad \mu_{Ps}=\frac{\hat q}2\ , \quad
(\tau_{SV})_{\mu}=-2q_{\mu} .
\end{equation}
In the other channels the four-quark condensates have more
complicated structure. In the vector and axial channels
\begin{eqnarray}
&& h^{V(A)}_{\mu\nu}\ =\ a^{V(A)}_hg_{\mu\nu} +  b^{V(A)}_h
  \frac{p_\mu p_\nu}{m^2}\ ,
\nn \\
\label{78} &&
r^{V(A)}_{\mu\nu}\ =\
a^{V(A)}_rg_{\mu\nu}+b^{V(A)}_r\frac{p_\mu p_\nu}{m^2}\ ..
\end{eqnarray}
Using Eqs. (\ref{73}), (\ref{74}) we obtain
\begin{equation} \label{79}
\mu_Vh^V=-a^V_h\hat q-b^V_h\frac{\hat p(pq)}{m^2}\ , \quad
\mu_Ah^A=a^A_h\hat q+b_h^A\frac{\hat p(pq)}{m^2}
\end{equation}
and
\begin{eqnarray}
\tau_Vr^V &=&\left(-10a^V_r-2b^V_r\right)\hat q-2b_r^V \frac{\hat
p(pq)}{m^2}\ , \nn \\
\label{80}
\tau_Ar^A &=&\left(-6a^A_r-2b^A_r\right)\hat q+2b_r^A \frac{\hat
p(pq)}{m^2}\ .
\end{eqnarray}
In the tensor channel
\begin{equation}
h^T_{\mu\nu,\rho\tau}\ =\
a^T_hs_{\mu\nu,\rho\tau}+b^T_ht_{\mu\nu,\rho\tau}
\label{81}
\end{equation}
with
\begin{eqnarray}
s_{\mu\nu,\rho\tau} &=& g_{\mu\rho}g_{\nu\tau}-g_{\mu\tau}g_{\nu\rho}\ ,
\nn \\
t_{\mu\nu,\rho\tau} &=& \frac1{m^2}\bigg(p_\mu p_\rho g_{\nu\tau} +
p_\nu p_\tau g_{\mu\rho}-p_\mu p_\tau g_{\nu\rho}-p_\nu p_\rho
g_{\mu\tau} \bigg)
\label{82}
\end{eqnarray}
and
\begin{equation}
\label{83}
\mu_T h^T\ =\ b^T_h\left(-\frac{\hat q}2+\frac{2\hat
p(pq)}{m^2}\right).
\end{equation}

The complete set of the four-quark expectation values
$a^X_r, b^X_r, a^X_h, b^X_h$ was
obtained in \cite{20} by using the approach motivated by the
perturbative chiral quark model (PCQM)
\cite{21,22}.
As explained in Introduction, the valence quarks are treated as the
relativistic constituent quarks, while the sea quarks are approximated
by those of perturbatively treated pions.

There are three types of contributions to the expectation values in the
approach of \cite{20}. All four operators can act on the constituent
quarks. Also, four operators can act on the pions. There are also the
"interference terms" with two of the operators acting on the valence
quarks while the other two act on the pions.

The contribution, corresponding to all four operators acting on pions
is expressed in terms of the pion expectation values of the four-quark
operators. The distribution of the pion field is determined
by the PCQM.
The contribution is
\begin{equation} \label{84}
(\Pi_{4q})_{pions}=\frac1{16q^2}\left(\sum_{X,\alpha}\langle \pi^\alpha
|\mu_X\bar u\Gamma^Xu\bar u\Gamma^Xu+4\tau_X\bar d\Gamma^Xd\bar
u\Gamma^Xu|\pi^\alpha\rangle\right)\frac{\partial\Sigma}{\partial
m^2_\pi}
\end{equation}
with $\Sigma$ standing for the sum of the self-energy and pion-exchange
contributions, while $"\alpha"$ denotes the
pion isotopic states. Using the values of the four-quark operators
averaged over pions \cite{23}, we find that
\begin{equation}
\label{85}
\sum_{X,\alpha}\mu_X\langle\pi^\alpha|\bar u\Gamma^Xu \bar
u\Gamma^Xu|\pi^\alpha\rangle+4\sum_{X,\alpha}\tau_X\langle\pi^\alpha|
\bar d\Gamma^Xd\bar u\Gamma^Xu|\pi^\alpha\rangle=0\ .
\end{equation}

Due to Eq. (\ref{85}) we can omit the contributions to the second
terms of the rhs of Eqs. (\ref{73}) and (\ref{74}) which are caused by
the pions only. Since the terms $\langle0|\bar qq|0\rangle
\langle\pi|\bar qq|\pi\rangle$ emerge as the ingredients of the
expectation values $\langle\pi|\bar qq\bar qq|\pi\rangle$ \cite{23},
the cancellation (\ref{85}) influences the first terms of rhs of Eqs.
(\ref{73}) and (\ref{74}) as well. Thus, in order to calculate the rhs
of Eq. (\ref{75}) it is sufficient to substitute for the operators with
the same flavor
\begin{equation}
\label{86}
h^X=2\cdot\frac56\langle 0|\bar u\Gamma^Xu|0\rangle\langle N|(\bar
u\Gamma^Xu)_v|N\rangle +\langle N|(\bar u\Gamma^Xu\bar u\Gamma^Xu)_1
|N\rangle\ .
\end{equation}
Here the lower index $"v"$ means that the operators act on the valence
quarks only. The lower index $"1"$ corresponds to the sum of the term
in which all the four operators act on the valence quarks and the term
in which two of the operators act on the valence quarks while the other
two act on pions. Of course, the first term in the rhs of Eq.
(\ref{86}) obtains a nonvanishing value only in the scalar case
$\Gamma^X=I$.

The expectation values of the operators of different flavors, providing
nonvanishing contributions to the rhs of Eq. (\ref{74}) are the
scalar-vector condensate
\begin{equation}
\label{87}
r^{SV}_\mu=2\cdot\frac23\langle0|\bar dd|0\rangle\langle N|\bar
u\gamma_\mu u|N\rangle+\langle N|(\bar dd\bar u\gamma_\mu u)_1|N\rangle
\end{equation}
and
\begin{equation}
\label{88}
r^X_{\mu\nu}\ =\ \langle N|\left(\bar d\Gamma^X_\mu d\bar u\Gamma^X_\nu
u\right)_1|N\rangle
\end{equation}
with $X$ standing for vector or axial structures. In the first
term in the rhs of Eq.(\ref{87}) the nonlocality of the vector
condensate is included.

The meaning of the
lower index $"1"$ is the same as in Eq. (\ref{86}).

Using the complete set of the nucleon four-quark expectation values
\cite{20}, we obtain
\begin{equation}\label{90}
(\Pi)_{4q}\ =\ \left(A^q_{4q}\frac{\hat q}{q^2}+A^p_{4q}
\frac{(pq)}{m^2}\frac{\hat
p}{q^2}+A^I_{4q} m\frac{I}{q^2} \right)\frac{a}{(2\pi)^2}\rho
\end{equation}
with the coefficients
\begin{equation}\label{91}
A^q_{4q}=0.25\,\quad
A^p_{4q}=-0.57\,\quad
A^I_{4q}=1.90\,
\end{equation}
and with the conventional notation
\begin{equation}\label{91a}
a = -(2\pi)^2\langle 0|\bar uu|0\rangle .
\end{equation}
We use the value $\langle 0|\bar uu|0\rangle$=(-241 MeV)$^3$,
corresponding to $a$=0.55 GeV$^3$, employed in \cite{6}. Note that $a$
is just a convenient scale for presentation of the results. It does not
reflect the chiral properties of $\Pi_{4q}$.

We can trace the structure of the three terms,
composing $\Pi_{4q}$ determined by Eq.(\ref{90})-see Appendix C.
The $\hat q$ term results mainly as the sum of the expectation value of
the product of the four $u$-quark operators, described by the first
(factorized) term
on the rhs of Eq. (\ref{86}), and that of the product of two $u$
and two $d$-quark operators in  the axial channel-Eq. (\ref{88}).
The $\hat p$ term is determined mostly by the expectation value
(\ref{88}) in the vector channel. The contribution proportional to
the unit matrix $I$ is determined by the scalar-vector expectation
value (\ref{87}). It is dominated by the first (factorized) term on
the rhs, while the second term diminished the value by about
30$\%$.

The contributions of the four-quark condensates to the lhs of the Borel
transformed sum rules (\ref{45})--(\ref{47})  are
\begin{eqnarray}
&& \omega^i=\omega^i_N\rho\ ; \quad \omega^i_N=A^i_{4q}f^i_{4q}\ ;
 \nn \\
&&  f^q_{4q}=-8\pi^2 a ,
\quad f^p_{4q}=-8\pi^2\frac{s-m^2}{2m} a ;
\quad f^I_{4q}=-8\pi^2 m a.
\label{92}
\end{eqnarray}

Note that we can modify our model approach by employing a more
sophisticated model for the pion. Namely, among the interference terms
contributing to the four-quark condensates, there is so-called "vertex
interference", in which one of the  vertices of the self-energy of
the valence quark is replaced by the four-quark operator. Some of such
terms contain the matrix elements $\langle0|\bar
u\gamma_5d|\pi^-\rangle$ and $\langle0|\bar d\gamma_5u|\pi^+\rangle$,
contributing to the expectation values $\langle N|\bar u\gamma_5d\bar
d\gamma_5u|N\rangle$, being connected with the matrix elements
$\langle N|\bar d\Gamma^Xd\bar u\Gamma^Xu|N\rangle$ of all structures
$\Gamma^X$ by the Fierz transform. On the other hand, they depend on
the values of the quark masses, since $\langle0|\bar
u\gamma_5d|\pi^-\rangle=-\frac{i\sqrt2\,F_\pi M^2_\pi}{m_u+m_d}$ with
$M_\pi$ ($F_\pi$) denoting the mass (decay constant) of pion. In a
somewhat straightforward approach one substitutes the current quark
masses.  Following more sophisticated models of the pions \cite{27}
one should substitute the constituent quark masses, thus obtaining much
smaller values. In the latter approach
\begin{equation}  \label{94}
A^q_{4q}\ =\ -\ 0.11,
\end{equation}
while the values of $A^p_{4q}$ and $A^I_{4q}$ remain unchanged.
In this case we find the larger cancellation between the first term of
rhs of Eq.
(\ref{86}) and the contribution coming from rhs of Eq. (\ref{88}).
 The latter is dominated by the vector expectation values.

\section{Sum rules in nuclear matter}

Actually, we shall solve the sum rules for the difference of the
operators in nuclear matter and in vacuum:
\begin{eqnarray}
\label{94'}
&& L^q(M^2,W^2_m,W^2_0)\ =\ \Lambda_m(M^2)-\Lambda_0(M^2)\ ;\\
\label{95}
&& L^p(M^2,W^2_m)\ =\ -\Sigma_v \Lambda_m(M^2)\ ;\\
\label{96}
&& L^I(M^2,W^2_m,W^2_0)\ =\ m^*\Lambda_m(M^2)-m\Lambda_0(M^2)
\end{eqnarray}
with $L^i(M^2,W^2_m,W^2_0)=L^i_m(M^2,W^2_m)-L^i_0(M^2,W^2_0)$. The
ingredients of Eqs. (\ref{94'})--(\ref{96}) are defined by Eqs. (25),
(26), (\ref{45})--(\ref{47}) and (\ref{50}).

Note that we took into account the anomalous dimensions only for
the leading  OPE terms $q^2\ln q^2$ and $\ln q^2$, neglecting
the anomalous dimensions of the $1/q^2$ OPE terms.

Although the anomalous dimensions of the
four-quark condensates are known \cite{28}, the anomalous dimension
matrix is not diagonal in the basis determined by Eq. (\ref{70}).
The calculation of this matrix in our basis is a separate work
which will be presented in further publications.
We use the nucleon structure functions presented in \cite{29},
which include the anomalous dimensions of the structure functions.

We solve Eqs. (\ref{94'})--(\ref{96}) in the same interval of the
values $M^2$ as it has been done in vacuum Eq. (33).

Since Eqs. (\ref{94'})--(\ref{96}) are not linear, the behavior of
the in-medium parameters is not linear in $\rho$ even if we limit
ourselves to the gas approximation. However, if the density $\rho$ is
small enough, we can try the linear approximation, assuming the
linear dependence of the nucleon characteristics on the density of
matter.

\section{Sum rules in the linear approximation}

It is instructive to express the density in units of the observable
saturation density
\begin{equation}
\label{97}
\rho_0\ =\ 0.17\, \rm fm^{-3}\ =\ 1.3\cdot10^{-3}\,GeV^3\ .
\end{equation}
The parameters which will be determined from the sum rules can be
presented as
\begin{eqnarray}
\Sigma_v=a_v\,\frac\rho{\rho_0}\ ; &&\ m^*=m+a_s\,
\frac\rho{\rho_0}\ ; \quad \delta\lambda^2=\lambda^{*2}_m-
\lambda^2_0=a_\lambda\frac\rho{\rho_0}\ ;
\nn \\
\label{98}
&& \delta W^2\ =\ W^2_m-W^2_0\ =\ a_W\,\frac\rho{\rho_0}\ .
\end{eqnarray}
To obtain the parameters in the linear approximation we put $Z=1$ and
find
\begin{eqnarray}
\Sigma_v &=& \Sigma_p\ , \quad m^*=m(1+\Sigma_q)+\Sigma_s\ ,
\quad  \lambda^{*2}_m=\lambda^2_m\ ,
\nn \\
m_m &=& m(1+\Sigma_q)+\Sigma_v+\Sigma_s\ =\ m^*+\Sigma_v
\label{99}
\end{eqnarray}
in Eqs. (38)--(40) and (\ref{49}). We put $s=4m^2$
in Eq. (39).

Expansion of the lhs of Eqs. (\ref{94'})--(\ref{96}) provides the
equations
\begin{eqnarray}
&& \left(f^q_v(M^2,W^2_0)v_N+f^q_g(M^2,W^2_0)g_N+u^q_N(M^2)
+w^q_N\right)\rho_0
\nn \\
&& \quad=\left  (a_\lambda-(a_s+a_v)\frac{2m\lambda^2_0}{M^2}\right)
e^{-m^2/M^2}-a_W\frac{\partial\ell^q_0(M^2,W^2_0)}{\partial W^2_0}\ ;
\label{100}
\\
&&\left(f^p_v(M,W^2_0)v_N+mu^p_N(M^2)+m\omega^p_N\right)\rho_0=\ -a_v
\lambda^2_0 e^{-m^2/M^2}\ ;
\label{101}
\\
&& \left(f^I_\kappa(M^2,W^2_0)\kappa_N+\omega^I_N(M^2)\right)\rho_0 =
\left(a_\lambda m-(a_s+a_v)\frac{2 m^2\lambda^2_0}{M^2}\right)
e^{-m^2/M^2}
\nn \\
&& \quad\ + a_s\lambda^2_0 e^{-m^2/M^2} -
a_W \frac{\partial\ell^I_0(M^2,W^2_0)}{\partial W^2_0}\ .
\label{102}
\end{eqnarray}
Note that in this form all three equations are tied. One can build
up another set of equations with the functions $L^p$, $L^I-mL^q$
and $L^q$ as the lhs. In this case the unknowns $a_v$ and $a_s$ are
determined from the separate equations. The third equation determines
the values of $a_\lambda$ and $a_W$. We introduce
\begin{eqnarray}
&& T^i_k(M^2,W^2_0)\ =\ \rho_0f^i_k(M^2,W^2_0)\frac{e^{m^2/M^2}}{
\lambda^2_0}\ (k=v,g,\kappa)\ ,
\nn \\
&& T^i_u(M^2,W^2_0)\ =\ \rho_0u^i_N(M^2,W^2_0)
\frac{e^{m^2/M^2}}{\lambda^2_0}\ ,
\nn \\
&& T^i_\omega(M^2)\ =\ \rho_0f^i_{4q}\ \frac{e^{m^2/M^2}}{
\lambda^2_0}
\label{103}
\end{eqnarray}
with the functions $f^i_k$ and $f^i_{4q}$ defined by Eqs.(\ref{54})
 and (\ref{92}).  We present
\begin{eqnarray} &&
T^p_v(M^2,W^2_0)v_N+mT^p_u(M^2,W^2_0)+mT^p_\omega A^p_{4q} =\ -\,a_v\;
\label{104} \\
&& T^I_\kappa(M^2,W^2_0)\kappa_N-mT^q_v(M^2,W^2_0)v_N-mT^q_g
(M^2,W^2_0)g_N-mT^q_u(M^2,W^2_0)
\nn \\
&& \quad+ T^I_\omega(M^2)A^I_{4q}-mT^q_\omega(M^2)A^q_{4q}\ =\ a_s;
\label{105}
\\
&& T^q_v(M^2,W^2_0)v_N+T^q_g(M^2,W^2_0)g_N+T^q_u(M^2,W^2_0)
+T^q_\omega(M^2)A^q_{4q}
\nn \\
&&\quad+ (a_s+a_v)\cdot\frac{2m}{M^2}\ =\
a_\lambda\frac1{\lambda_0^2} -a_W\frac1{\Lambda_0}
\frac{\partial\ell^q_0(M^2,W^2_0)}{\partial W^2_0}
\label{106}
\end{eqnarray}
with $\Lambda_0$ being defined by Eq. (27). The characteristics
$a_v$ and $a_s$ are found from Eqs. (\ref{104}) and (\ref{105}) and
are substituted into Eq. (\ref{106}). The latter determines the
values of $a_\lambda$ and $a_W$.

Note that there is one more approximation in the rhs of Eq.
(\ref{104}). Namely, we neglected the value
$$
a_W\left(\frac{\partial\ell^I_0}{\partial W^2_0}-m\,
\frac{\partial\ell^q_0}{\partial W^2_0}\right)=\ a_W W^2_0\left(
-2a+\frac{mW^2_0}{2L^{4/9}}+\frac{mb}{4W^2_0L^{4/9}}\right)e^{-W^2_0/M^2}
$$
since there is about 80$\%$ cancellation between the two terms on the
lhs in the interval determined by Eq.(33). This is due to the positive
 parity of the nucleon state --- See Appendix D.

The values of the QCD parameters which enter the lhs of Eq.
(\ref{104})--(\ref{106}) are determined by Eqs. (6'), (8), (10),
(65) and (86). The expectation value  $\kappa_N=\langle N|\bar uu+\bar
dd|N\rangle$ is connected to the pion-nucleon sigma term
$\sigma_{\pi N}$ by Eq.(8). The value of $\sigma_{\pi N}$  can be
extracted from the data on low-energy $\pi N$ scattering, being
expressed through the observable $\Sigma$ term ($\Sigma_{\pi N}$)
\cite{30}. The value
\begin{equation}
\label{107}
\sigma_{\pi N}\ =\ (45\pm7)\ \rm MeV\
\end{equation}
corresponds to $\Sigma_{\pi N}$=64 MeV \cite{31}.
We shall present the specific values, corresponding to $\kappa_N=8$.
This value corresponds to $\sigma_{\pi N}=45\,$MeV and the sum of the
light quark masses $m_u+m_d=11\,$MeV. There is an uncertainty in the
value of $\kappa_N$ due to the errors in determination of the values of
$\sigma_{\pi N}$ and $m_u+m_d$. We  also present the dependence of the
characteristics of the nucleon on the value of $\kappa_N$.

The values of the parameters
\begin{equation}
\label{108}
a_v=0.108\mbox{ GeV },\ a_s=-0.178\mbox{ GeV },\
a_\lambda=-1.29\mbox{ GeV}^6,\ a_W=-0.81\mbox{ GeV}^2
\end{equation}
are obtained by minimization of the relative difference of the rhs and
lhs of Eqs. (\ref{104})--(\ref{106}) by the chi-square method. The
solution of these equations is illustrated in Fig. 1.
Note, that if we construct the equation which is the difference of
Eqs.(\ref{106}), (\ref{107}) the function of $M^2$ in the lhs should be
approximated by the constant value $a_s+a_v$, having the meaning of
the potential energy. Such approximation holds with much better
accuracy then the separate Eqs.(\ref{106}) and (\ref{107}) for the
self-energies. The solution (\ref{108}) corresponds to the values
\begin{equation} \label{109}
\Sigma_v=108\mbox{ MeV }, \quad
m^*-m=-178\mbox{ MeV }, \quad
\frac{\delta\lambda^2}{\lambda^2_0}=-0.67\ , \quad \frac{\delta
W^2}{W^2_0}=-0.37\ .
\end{equation}
Although the sets of Eqs.(\ref{100})--(\ref{102}) and
(\ref{104})--(\ref{106}) are mathematically identical, a procedure of
matching of the two sides of the equations may lead to somewhat
different solutions. Applying the same procedure of minimization to the
set of Equations (\ref{100})--(\ref{102}) we find
$a_v=0.108\,$GeV $a_s=-0.254\,$GeV, $a_\lambda=-1.61\,\rm
GeV^6$, $a_W=-0.91\,\rm GeV^2$.  Thus the parameters
$\delta\lambda^2$ and $\delta W^2$ are determined with somewhat
larger uncertainties than the self-energy $\Sigma_v$.

As   we noted at the end of Sec. IV, our model approach to the
calculation of the four-quark condensates can be modified by using more
sophisticated models of the pions \cite{27}, \ie\ by the account of the
constituent quark masses. Using the value of $A^q_{4q}$ given by Eq.
(\ref{94}), we obtain from Eqs. (\ref{104})--(\ref{106})
\begin{equation} \label{110}
\Sigma_v=108\mbox{ MeV }, \quad m^*-m=-203\mbox{ MeV }, \quad
\frac{\delta\lambda^2}{\lambda^2_0}=-0.71\ , \quad \frac{\delta
W^2}{W^2_0}=-0.41
\end{equation}
at the saturation density $\rho=\rho_0$.
Thus, this change of the value $\omega^q_N$ results in
the change of the nucleon parameters by less than $15\%$.

Note that the functions $T^i_j(M^2)$ defined by Eq. (\ref{103})
$(j=v,g,\kappa,u,\omega; i=q,p,I)$ depend on $M^2$ rather weakly.
Thus, approximating
\begin{equation} \label{111}
T^i_j(M^2)\ =\ C^i_j\ ,
\end{equation}
we can replace in the lhs of Eqs. (\ref{104})--(\ref{106}) the functions
$T^i_j(M^2)$ by the constant coefficients $C^i_j$.
Numerically the most important functions $T^p_v(M^2)$ and
$T^I_\kappa(M^2)$ can be approximated by the constant values with the
errors of about 4\% and 8\%.  The largest errors of about $25\%$
emerge in the averaging of the  functions $T^i_\omega$. This
solves the problem of expressing the in-medium change of nucleon
parameters through the values of the condensates
\begin{eqnarray} \label{112}
\Sigma_v &=& -\left(C^p_v v_N+mC^p_u+mC^p_\omega
A^p_{4q}\right)\frac\rho{\rho_0}\ ; \\
m^*-m &=& \left(C^I_\kappa\kappa_N-mC^q_vv_N-mC^q_gg_N-mC^q_u +
C^I_\omega
A^I_{4q}-mC^q_\omega A^q_{4q}\right)\frac\rho{\rho_0}\ .
\label{113}
\end{eqnarray}
The coefficients in the rhs of Eqs. (\ref{112}) and (\ref{113}) are
\begin{eqnarray}
&&
C^q_v=-0.062, \quad
C^q_g=0.011\mbox{ GeV}^{-1}, \quad
C^q_\omega=-0.067, \quad
C^q_u=-0.074,
\nn \\
&&
C^I_\kappa=-0.042\mbox{ GeV },\quad
C^I_\omega=-0.064\mbox{ GeV },
\nn \\
&&
C^p_v=-0.090\mbox{ GeV }, \quad
C^p_\omega=-0.095\ ,\quad
C^p_u=0.094\ .
\label{114}
\end{eqnarray}
Equations (\ref{112}) and (\ref{113}) reproduce the values of
$\Sigma_v$ and $m^*$ provided by Eqs. (\ref{109}) with the accuracy of
15$\%$ and 6$\%$ correspondingly.\\

\section{Beyond the linear approximation}

Now we do not assume the linear dependence of the nucleon parameters on
the density $\rho$. We find the values $\Sigma_v$, $m^*$,
$\lambda^{*2}_m$ and $W^2_m$ which minimize the difference between lhs
and rhs of Eqs. (\ref{94'})--(\ref{96}). The consistency of the
lhs and rhs is illustrated by Fig. 2.
At the saturation density $\rho=\rho_0$ we obtain
\begin{equation} \label{115}
\Sigma_v=150\mbox{ MeV }, \quad m^*-m=-200\mbox{ MeV }, \quad
\lambda^{*2}_m=1.25\mbox{ GeV}^6, \quad W^2_m=2.11\mbox{ GeV}^2.
\end{equation}
The two last numbers correspond to the relative shifts
$\delta\lambda^2/\lambda^2_0=-0.35$ and $\delta W^2/W^2_0=-0.03$.
Thus the linear approximation is true at $\rho\approx\rho_0$ with
the accuracy of about $25\%$ for the vector self-energy and
about 10$\%$ for the scalar one.
The linear approximation overestimates the shift of the
effective threshold.

Recall, that we presented the numerical results for $\kappa_N=8$.  The
dependence on the value of $\kappa_N$ is shown in Fig.  3.
The density dependence of the nucleon parameters at $\kappa_N=8$ is
shown in Fig. 4.

Using Eq.
(\ref{94}) for the value of $A^q_{4q}$ we obtain the results which are
close to those presented by Eq.  (\ref{115})
\begin{equation} \label{116}
\Sigma_v=142\mbox{ MeV }, \quad m^*-m=-223\mbox{ MeV },
\quad \lambda^{*2}_m=1.24\mbox{ GeV}^6, \quad W^2_m=2.09\mbox{ GeV}^2.
\end{equation}

Note that the difference between the linear and nonlinear solutions
has a strong effect on
 the value of the nucleon potential energy
\begin{equation} \label{117}
U(\rho)\ =\ \Sigma_v(\rho)+m^*(\rho)-m\ ,
\end{equation}
which is about -40 MeV and -70 MeV for the solutions (\ref{115})
and (\ref{116}) at the phenomenological saturation point
$\rho=\rho_0$ .

\section{Discussion}

It is instructive to follow how the values of the nucleon self-energies
change, while we include the various contributions of  the lhs of the
sum rules. The solutions of the general equations (92)-(94) are
illustrated by Fig. 5. At the saturation
density $\rho_0$ the vector self-energy $\Sigma_v$ and the effective
mass $m^*$ are 340~MeV and 750~MeV correspondingly, if only
the terms $l^j$ -Eq. (\ref{53}) are included
 in $L^i_j$ of Eq.(\ref{50}). One can see from
Fig. 5 that the higher moments of the structure functions and the
four-quark condensates subtract about
100~Mev each from the value of $\Sigma_v$.
On the contrary, the two contributions to $m^*$ cancel to large extent,
with the four-quark condensate subtracting about 200~MeV, and the
moments of the structure functions adding about this value.

We come
to similar results in the linear approximation Sec. VI. The
moments of the structure functions and the four-quark condensate
subtract 60~MeV and 110~MeV from the lowest dimension value
$\Sigma_v$=270~MeV.
The OPE value of the scalar self-energy $m^*-m$ is
 -140~MeV. The four-quark condensates and the moments of the structure
functions add -140~MeV and +100~MeV, correspondingly.

Turning to the role of the anomalous dimensions, we note that their
inclusion into the moments of the structure functions lead to minor
changes of several MeV of the values of vector and scalar self
energies.  Neglecting the anomalous dimensions of all the
in-medium contributions increases the values of the vector
self-energy $\Sigma_v=230$~MeV, and  of the scalar self-energy
$m^*-m = -140$~MeV. Thus, we find for the potential energy $U>0$
if $\kappa=8$. However, the value of $m^*$ decreases with
$\kappa$, while the vector self energy practically does not
change. We find that $U<0$ if $\kappa>10$, i.e.
$\sigma_{\pi N}>55$~MeV - Eq.(\ref{107}).

The authors of \cite{16} carried out the detailed analysis of the
nucleon self-energies depending on the in-medium values of the
four-quark condensates.  They considered the QCD sum rules, based on
the dispersion relations in energy $q_0$ at $|{\bf q}|$ being fixed.
The authors of \cite{16}
found that the values of the self-energies depend strongly on
the value of scalar-scalar condensate, while the dependence on
the values of the other four-quark expectation values appeared
to be negligible small. Actually, they presented
$\langle M|\bar qq\bar qq|M\rangle-\langle 0|\bar qq\bar qq|0\rangle =
  2 f\langle 0|\bar qq|0\rangle \langle N|\bar qq|N\rangle\rho$, and
studied the dependence of the nucleon parameters on the value of
$f$.  Our model calculations correspond to $f=0.14$. It was
found in \cite{16}, that the values $0<f<0.3$ provide the
results, which are consistent with the nuclear phenomenology.
We can deduce from Fig.1 of \cite{16}, that there values are
 $m^*/m=0.65$ and $\Sigma_v/m=0.28$ for $f=0.14$. Neglecting
 all the other four-quark condensates, we find the close values
$m^*/m=0.67$ and $\Sigma_v/m=0.25$. Note, however, that our
approach is based on the dispersion relations in another
variable, \ie\ in $q^2$, with the
relativistic pair energy $s$ being kept fixed. (This enables us
to avoid the singularities, connected with the excitations of
medium \cite{8}-\cite{10}.)  In our case the influence of the
vector-scalar expectation value is stronger, then in \cite{16}.
For example, if we assume factorization approximation for the
vector-scalar condensate, our value of the nucleon effective
mass is about twice smaller, then the value, obtained in
\cite{16}. The values of the vector self-energy are still close
in the two approaches.

In the paper \cite{17} the calculations of the four-quark condensate
were avoided by a specific choice of the function $\Pi_m(q)$.
The limits 160~MeV$<\Sigma_v<$310~MeV and
 0.62~GeV$<m^*<$0.83~GeV have been obtained.
 In the work \cite{9} the authors
got rid of the four-quark condensates, applying the differential
operators. They found the vector and scalar fields to be about
220~MeV and -350~MeV in the gas approximation.

These results are consistent with each other and with the results of
nuclear physics calculations. Various approaches in the nuclear physics
studies ( see, e.g. \cite{32}) provide the values between 180~MeV
and 350~MeV for the vector fields, and between -200~MeV and
-400~MeV for the scalar fields.

There is agreement with the  earlier results in some other
points. The $30\%$ reduction of the vector field, caused by nonlocality
of the vector condensate, was found in \cite{8,9}. The strong reduction
of the nucleon pole residue was obtained in \cite{8,9,16}. Also,
it was first noted in \cite{16} that the shift
of the continuum threshold is very small.

\section{Summary}

We analyzed QCD sum rules in nuclear matter
by taking into  account
terms of the order of $q^2\ln q^2$, $\ln q^2$
and $1/q^2$ of the operator product
expansion. The consistency of the lowest OPE terms
in QCD sum rules
\cite{7,8,9,10,14,15,16} with the nuclear phenomenology was known for
a long time. However the lack of information on the  four-quark
condensates, contributing to the terms of the order $1/q^2$ was the
main obstacle for the further development of the approach.

In this paper we studied the sum rules, treating the QCD
condensates in the gas approximation and included the
contribution of the four-quark condensates, expressed through
the nucleon expectation values. The latter were obtained in
\cite{20} by employing results of  the perturbative chiral quark
model \cite{21,22}. We included also the higher moments of the
nucleon structure functions which contribute to the terms of the
order $\ln q^2$ and $1/q^2$. Taking into account the four-quark
condensate we included all Lorentz structures.

We took into account the nonlocal structure of the vector condensate,
which manifests itself through the higher moments of the structure
functions. We include corrections, which have the smooth dependence on
these moments. However, we did not include the nonlocal singularities
in the $u$-channel. Such singularities correspond to the exchange
interaction between the nucleon and the matter. Thus, our approach
corresponds to Hartree description of the in-medium nucleon.
The nonlocal structure of the scalar condensate manifests
itself in the higher orders of OPE.

Considering only the linear changes of the nucleon parameters we
obtained a linear combination of the QCD sum rules equations in which
the nucleon effective mass $m^*$ and the vector self-energy
$\Sigma_v$ are the only unknown parameters. A more detailed analysis
going beyond the linear approximation shows that this approach works
well at  densities close to the saturation value $\rho=\rho_0$. In
this approach we solved the problem of expressing the in-medium
change of the nucleon parameters in terms of the in-medium values of
QCD condensates --- Eqs.  (\ref{112}) and (\ref{113}).

The terms, containing the four-quark condensates provide
the corrections of the order $20-25\%$ to the leading terms of
the OPE of the function $\Pi_m-\Pi_0$, which are determined
by the local vector and scalar condensates.  This is
consistent with the hypothesis about the convergence of the
OPE series.  The four-quark condensates diminish the
OPE value of the vector self -energy $\Sigma_v$ by about $25\%$.
The scalar self-energy $m^*-m$ is more sensitive to the
four-quark expectation values. Inclusion of these
condensates makes the OPE value of $m^*-m$ about $80\%$ larger.
Inclusion of the nonlocality of the vector condensate, which manifests
itself in terms of the higher moments of the structure functions
subtracts $25\%$ more from the value of $\Sigma_v$, and almost
totally compensates the contribution of the four-quark
condensates to the shift $m^*-m$. Thus the value of $m^*-m$
appears to be very close to the one, determined by the lowest
orders of OPE.

The contribution of the four-quark condensate to the vector self-energy
$\Sigma_v$ is caused mainly by the vector-vector structure. The
contribution to the scalar parameter $m^*-m$ is of more complicated
origin, with the scalar--vector, scalar--scalar, vector--vector and
axial--axial terms being numerically important.

As it was noted earlier \cite{9,10}, the QCD sum rules can be viewed as
a connection between the exchange of uncorrelated $\bar qq$ pairs and
the exchange of strongly correlated pairs with the same quantum
numbers (mesons). This results in a certain connection between the
Lorentz structures of the in-medium expectation values and of the
nucleon propagator. In the leading orders of OPE the vector (scalar)
structure of the propagator is determined by the vector (scalar)
expectation value. The scalar-vector four-quark condensate is
determined mainly by the contribution which is proportional to the
vector expectation value.  On the other hand, it contributes to the
scalar Lorentz structure of the nucleon propagator. In the
meson-exchange picture such terms can be explained by the complicated
structure of the nucleon-meson vertices.  This can be instructive for
model building of  nuclear forces.

The values of the nucleon parameters $\Sigma_v$ and $m^*-m$ are
(at least qualitatively) consistent with those,
 obtained earlier in framework of nuclear physics \cite{32} and
of QCD sum rules approach \cite{7}-\cite{9}, \cite{14}-\cite{17}.
The four-quark condensates, as well as the higher
moments of the structure functions provide large contributions
to the nucleon parameters. This future accounting for the main
radiative corrections is expected to make the results more
accurate.

Another  direction of the development of the approach is to
go beyond the gas approximation.
The presentation
of the results, especially Eq.  (\ref{113}) for $m^*$ enables to
make the next step, studying the self-consistent set of equations
for the nucleon effective mass and the quark condensates, as
suggested in \cite{10}.

\section{Acknowledgments}

We thanks V.E.Lyubovitskij for discussions.
Two of us (E.G.D. and V.A.S.) are grateful for the hospitality of
University of T\"ubingen during their visit.
The work was supported by
the DFG under contract 438/RUS 113/595/0-1, FA67/25-3 and
GRK 683, and by the grants RFBR 03-02-17724 and RSGSS-1124.2003.2.

\section{Appendix A}
The contribution $X_{1a}(q)$ expressed by Eq. (\ref{63}) can be
obtained by direct substitution of Eq. (\ref{56}) into Eq.
(\ref{58}) and by using the formula
$$
\int\frac{d^4x}{x^8}\,(ax)(bx)e^{i(q'x)}=\frac16\left[(ab)+
\frac{2(aq')(bq')}{q^{'2}}\right]\int \frac{d^4x}{x^6}e^{i(q'x)}
\eqno{(A1)}
$$
for any vectors $"a"$ and $"b"$. Thus, all the contributions to the
function $X_{1a}$ are expressed through the integral
$$
\int \frac{d^4x}{x^6}\,e^{i(q'x)}\ =\ -\ \frac{i\pi^2}8
q^{'2}\ln(-q^{'2}) + \ldots\ .
  \eqno{(A2)}
$$
Here the dots denote the
terms which will be killed by the Borel transform. This leads to Eq.
(\ref{63}).

To establish the connection between Eq. (\ref{63}) and the two terms on
the rhs of Eq. (41) note, that the rhs of Eq. (\ref{63}) consists of
the terms of the form (see Eq. (\ref{62}))
$$
X\ =\int\limits^1_0 d\alpha\ln(Q^2+A^2(\alpha))f(\alpha)\
=
\ln Q^2\int\limits^1_0 d\alpha f(\alpha)\
+\int\limits^1_0 d\alpha\ln\frac{Q^2+A^2(\alpha)}{Q^2}\,f(\alpha) .
  \eqno{(A3)}
$$
The first term on the rhs contains the standard logarithmic factor
containing the cut, running to infinity. It is described by our
"pole+continuum" model in a usual way. The second term contains a
finite cut. Such terms need special treatment. The cut of the second
term describes the singularities in the  $u$-channel, caused by the
nonlocal structure of the vector condensate. They correspond to the
exchange terms on the rhs of the sum rules. We neglect such
contributions, thus coming to the Hartree description of the nucleon in
nuclear matter.

\section{Appendix B}

To obtain the coefficients of the first (factorized) terms in the rhs
of Eqs. (83) and (84), recall that we need the expectation values of the
color-antisymmetric operators
$$
T^{XY,f_1f_2}=\left(:\bar q\,^{f_1a}\Gamma^X\bar q\,^{f_1a'}\cdot\bar
q\,^{f_2b}\Gamma^Yq^{f_2b'}:\right)(\delta_{aa'}\delta_{bb'}
-\delta_{ab'}\delta_{ba'})
  \eqno{(B1)}
$$
with $f_1,f_2$ standing for the quark flavors. The dots denote the
normal ordering of the operators, $a,a',b,b'$ represent the color
indices. It is convenient to present
$$
\delta_{aa'}\delta_{bb'}-\delta_{ab'}\delta_{ba'}\ =\ \frac23\,
\delta_{aa'}\delta_{bb'}-\frac12\,\sum_\rho\lambda^\rho_{aa'}
\lambda^\rho_{bb'}
  \eqno{(B2)}
$$
with $\lambda^\rho$ standing for the SU(3) Gell-Mann matrices
Tr$\,\lambda^\rho\lambda^\sigma=2\delta^{\rho\sigma}$.

The factorization approximation for the quarks of different flavors is
$$
\langle M|\bar u_\alpha^au_\beta^{a'}\bar
d_\gamma^bd^{b'}_\delta|M\rangle\ =\ \langle M|\bar
u_\alpha^au^{a'}_\beta|M\rangle \langle M|\bar
d_\gamma^bd^{b'}_\delta|M\rangle
  \eqno{(B3)}
$$
with $\alpha,\beta, \gamma,\delta$ being the Lorentz indices, and only
the first term of the rhs of Eq. (B2) contributes. Using also Eq. (13)
we come to Eq. (71) and (84).

The factorization approximation formula for the quarks of the same
flavor, \eg\ $q^{f_1}=q^{f_2}=u$ is
$$
\langle M|\bar u\,^a_\alpha u^{a'}_\beta\bar u\,^b_\gamma
u^{b'}_\delta |M\rangle\ =\ \langle M|\bar u\,^a_\alpha u^{a'}_\beta|M
\rangle\langle M|\bar u^b_\gamma u^{b'}_\delta|M\rangle
\nn \\
-\ \langle M|\bar u\,^a_\alpha u^{b'}_\delta|M\rangle\langle
M|\bar u\,^b_\delta u^{a'}_\beta|M\rangle\ .
\eqno{(B4)}
$$
Thus in the factorization approximation
$$
\langle M|\bar u\Gamma^Xu\bar u\Gamma^Yu|M\rangle=\frac1{16}\left[
\tr\Gamma^X\cdot\tr\Gamma^Y-\frac13\tr(\Gamma^X\Gamma^Y)\right]
(\langle M|\bar uu|M\rangle)^2
  \eqno{(B5)}
$$
and
$$
\langle M|\sum_\rho\bar u\Gamma^X\lambda^\rho u\cdot\bar u\Gamma^Y
\lambda^\rho u|M\rangle\ =\ -\frac19\tr(\Gamma^X\Gamma^Y)(\langle
M|\bar uu|M\rangle)^2.
  \eqno{(B6)}
$$
Thus, for the factorized part of the expectation value of the
color-antisymmetric operator $T^{II,uu}$ is
$(\Gamma^X=\Gamma^Y=I)$ $$ \langle M|\bar u\,^au^{a'}\bar
u\,^bu^{b'}|M\rangle\ =\ C(\langle M|\bar uu|M\rangle)^2
  \eqno{(B7)}
$$
with
$$
C\ =\ \frac23\left(1-\frac1{12}\right)-\frac12\left(-\frac49\right)
=\ \frac56\ .
  \eqno{(B8)}
$$
Employing also Eq. (13) we come to Eq. (70) and (83).

\section{Appendix C}

Here we present for illustration the calculation of the most
important contributions of the four-quark condensates to $\hat q$
structure. Using Eq. (\ref{75}) we find for the contribution of the
first term of the rhs of Eq. (\ref{86})
$$
\Pi^1=\left(-\frac12\right)2\cdot\frac56\cdot\frac{\langle0|\bar uu|
0\rangle}{q^2}\left[\langle p|(\bar uu)_v|p\rangle\rho_p+\langle
n|(\bar uu)_v|n\rangle\rho_n\right].
  \eqno{(C1)}
$$
This is equivalent to
$$
\Pi^1\ =\ -\,\frac54\,\frac{\langle0|\bar uu|0\rangle}{q^2}\cdot
J\rho
  \eqno{(C2)}
$$
with $J=\int\bar\psi(x)\psi(x)d^3x$, while $\psi(x)$ is the
renormalized PCQM wave function of the constituent quark, normalized
by the condition $\int\bar\psi(x)\gamma_0\psi(x)d^3x=1$. Using the
value $J=0.54$ \cite{22}, one finds
$$
\Pi^1\ =\ -0.67\,\frac{\langle0|\bar uu|0\rangle}{q^2}\,\rho\ .
  \eqno{(C3)}
$$

The interval contribution is determined mostly by the expectation
values of the operators $\bar d\Gamma^Xd\bar u\Gamma^Xu$. This
happens just due to the large numerical coefficients on the rhs of
Eq. (\ref{78}). Using Eq. (\ref{78}) we find the contribution to be
$$
\Pi^2\ =\ \left(-10a^V_r-6a^A_r-2b^V_r-2b^A_r\right)\frac\rho{q^2}\ .
  \eqno{(C4)}
$$
Substituting the values $a^V_r=-0.074\varepsilon^3_0$,
$a^A_r=0.084\varepsilon^3_0$,  $b^V_r=0.31\varepsilon^3_0$,
$b^A_r=0.06\varepsilon^3_0$  ($\varepsilon_0=241$MeV)
\cite{20}, we obtain
$\Pi^2=-0.50(\varepsilon^3_0/q^2)$ and
$$ \Pi^1+\Pi^2\ =\ 0.17\ \frac{\varepsilon^3_0\rho}{q^2}\ .
  \eqno{(C5)}
$$

A more accurate calculation, accounting for the internal
contributions of the operators $\bar u\Gamma^Xu\bar u\Gamma^Xu$ leads
to the first term in the rhs of Eq. (\ref{91}).

The contribution to $\hat p$ structure is obtained in similar way.
Turning to $I$ structure, note that the contribution comes from the
scalar-vector condensate $\bar dd\bar u\gamma_\mu u$ --- Eq. (\ref{87}).
The first ("factorized") term in the rhs provides the contribution
$$
\Pi^3\ =\ -\frac23\int\frac{d^4x}{\pi^2 x^4}(x,\theta^u(x))
e^{i(qx)}\langle0|\bar dd|0\rangle\ \rho
  \eqno{(C6)}
$$
with $\theta^q(x)$ defined by Eq. (\ref{57}). If $\theta^u_\mu(x)=
\theta^u_\mu(0)$, we obtain
$$
\Pi^3\ =\ -\frac{2(pq)}{q^2}\ \langle0|\bar dd|0\rangle\ \frac\rho m.
$$
Taking into account the dependence of $\theta^u_\mu$ on $x$ we
actually include the higher moments and twists of the nucleon
structure functions. Proceeding in the same way as in Sec. III,
we obtain for the Borel transform of $\Pi^3$
$$
B\Pi^3\ =\ - 8\pi^2 m Y a \rho  .
  \eqno{(C7)}
$$
Here
$$
Y = \frac13\left(\frac{s-m^2}{m^2} \langle\eta\rangle
- \langle\alpha\eta\rangle
- \frac12\langle\xi\rangle
- \frac14\frac{m^2_0}{m^2}\langle\eta\rangle\right)
  \eqno{(C8)}
$$
The first term, that is the pure local contribution, would give
Y=3.0, the higher order contributions subtract 0.32 from this
value.  Thus, the factorized term would provide $A^I_{4q} =
2.68$.  Account of the second term on the rhs of (\ref{87})
leads to $A^I_{4q} = 1.90$.

\section{Appendix D}

Present vacuum sum rules given by Eqs. (29) and (30) in the form
$$
\ell^q_0(M^2,W^2_0)
= \Lambda_0+\int\limits_{W^2_0}
\frac{\partial\ell^q_0}{\partial W^2}\ dW^2\ ,
\nn \\
\ell^I_0(M^2,W^2_0)
= m\Lambda_0+\int\limits_{W^2_0}
\frac{\partial\ell^I_0}{\partial W^2}\ dW^2
  \eqno{(D1)}
$$

with $\Lambda_0(M^2)$ determined by Eq. (27). In the combination
$\ell^I_0-m\ell^q_0$ which is just the projection on the
negative-parity component of the lowest state the contribution of the
residue vanishes
$$
\ell^I_0-m\ell^q_0\ =\ \int\limits_{W^2_0}\left(\frac{\partial
\ell^I_0}{\partial W^2}-m\,\frac{\partial\ell^q_0}{\partial W^2}
\right)dW^2\ .
  \eqno{(D2)}
$$
The condition
$$
\left|\frac{\partial(\ell^I_0-m\ell^q_0)}{\partial W^2}\right|\
\ll\ \left|\frac{\partial\ell^{q,I}_0}{\partial W^2}\right|
$$
at $W^2=W^2_0$ means that we cannot imitate the contribution of the
negative-parity pole of the order $\Lambda^2_0$ in rhs of Eq. (D2)
by changing the value of $W^2_0$.

\newpage

\section{Figure captions}
\noindent
Fig.1.
Solution of Eqs. (\ref{104})--(\ref{106}). In Fig. 1a the lines
1 and 2 show the lhs of Eqs. (\ref{104}) and (\ref{105}) for the
self-energies. Line 3 shows the  potential
energy.  The dashed lines show the constant values,
corresponding to $a_v$ and $a_s$ in rhs of these equations. The
line in Fig. 1b shows the ratio of lhs and rhs of Eq.
(\ref{106}).

\noindent
Fig.2.
Curves 1, 2 and 3 show the lhs to rhs ratios of Eqs.
(\ref{94'})--(\ref{96}) correspondingly, at the values of the
nucleon and continuum parameters given by Eq. (\ref{115}).

\noindent
Fig.3.
Dependence of the solutions of Eqs. (\ref{94'})--(\ref{96}) on
the value of the nucleon expectation value $\kappa_N$ at
$\rho=\rho_0$.
The values of $W^2_0$, $\lambda^2_0$  are given by Eq.(33').

\noindent
Fig.4.
Density dependence of the nucleon and continuum parameters
beyond the linear approximation at $\kappa_N = 8$. The
horizontal axis corresponds to the density of the matter,
related to the phenomenological saturation value.

\noindent
Fig.5.
Sum-rule predictions for the dependence of the nucleon parameters
 $m^*/m$ and  $\Sigma_v/m$ on the ratio
 $\rho/\rho_0$ at $\kappa_N = 8$.
The  curves correspond to the successive inclusion of more complicated
condensates.
Dashed lines: only expectation values of the operators of the
lowest dimension $\bar q(0)\gamma_0 q(x=0)$ and $\bar
q(0)q(x=0)$ and of the gluon operators
$\frac{\alpha_s}{\pi}G^2(0)$ are included (see Eq.(\ref{53})).
Dotted lines: local four-quark condensates are added
(Eqs. (\ref{87}), (\ref{88})); solid lines:  $x$-dependence of
the vector condensates (expressed in terms of the nucleon
structure functions) is included.

%\end{document}

\newpage

\begin{figure}
\centering{\epsfig{figure=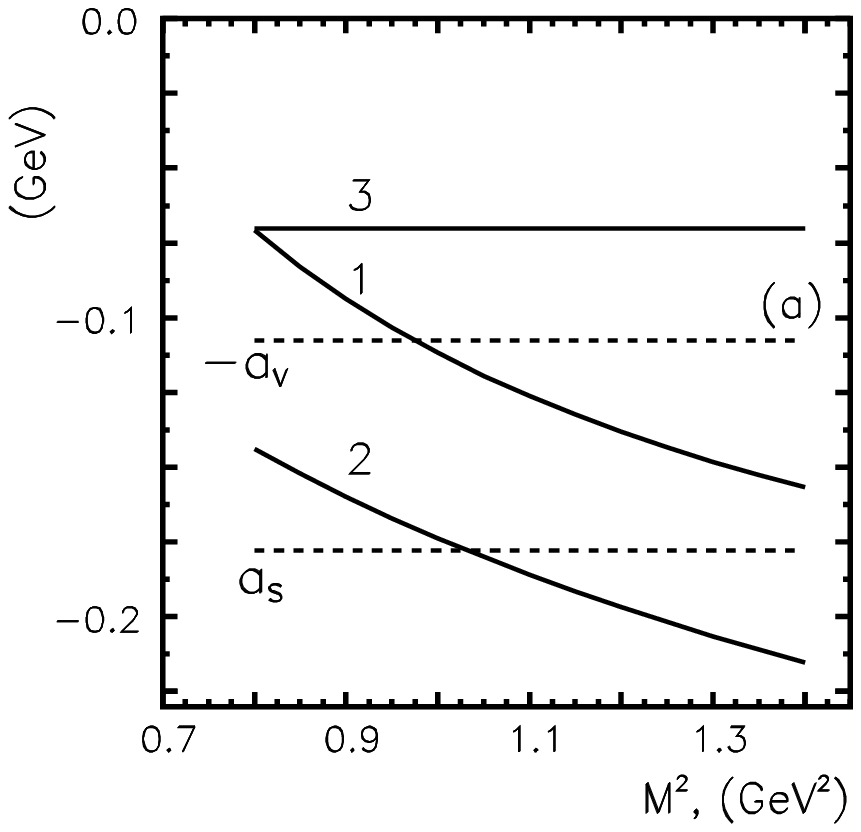,width=9cm}
 \epsfig{figure=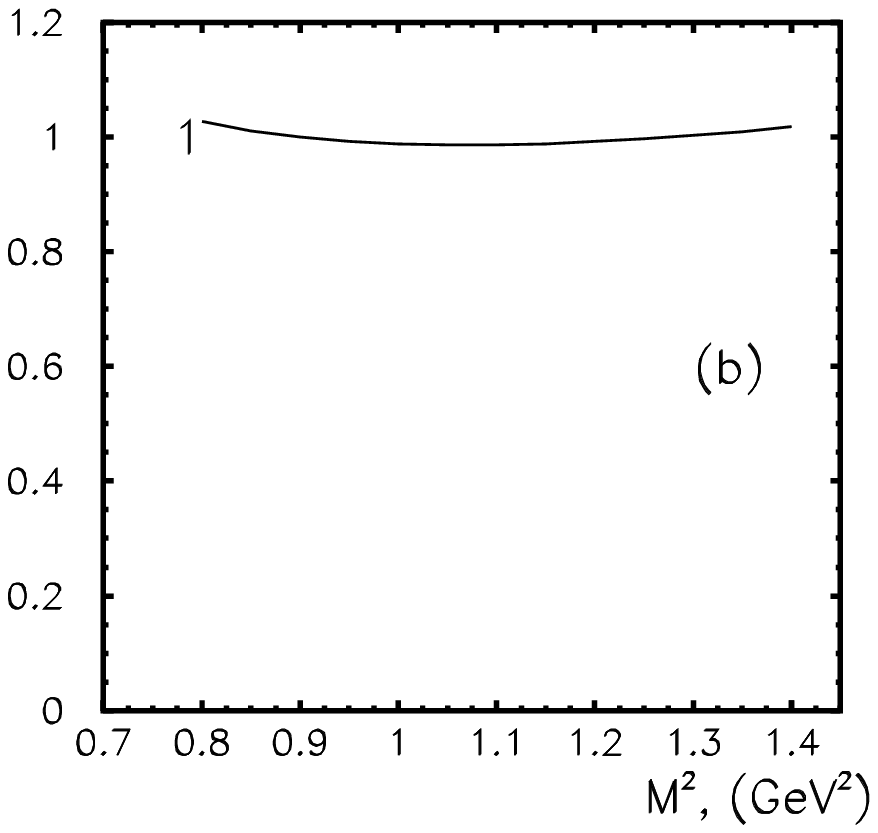,width=8.5cm}}
\caption{}
\end{figure}

\begin{figure}
\centering{\epsfig{figure=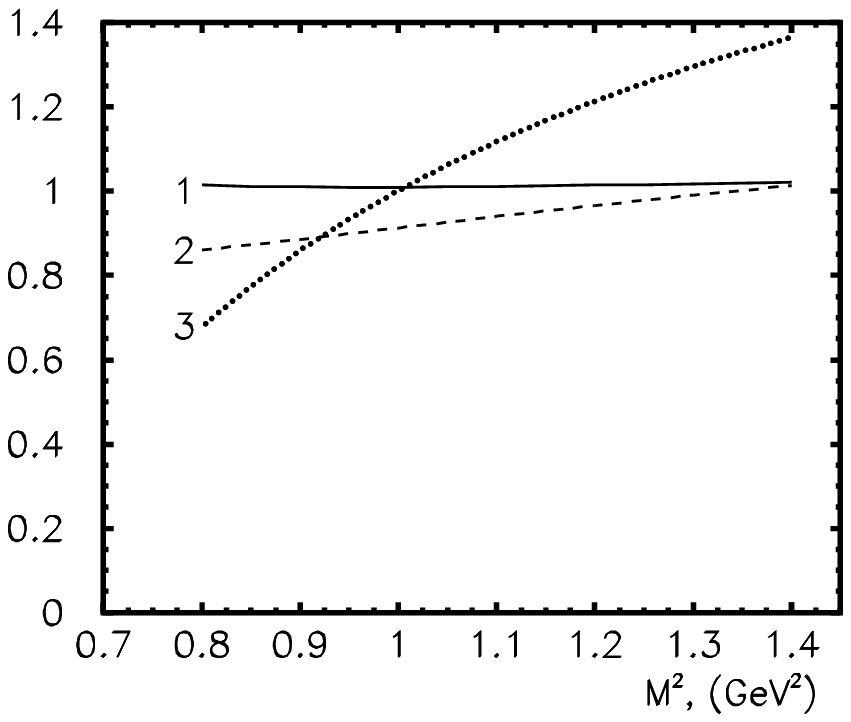,width=10cm}}
\caption{}
\end{figure}

\begin{figure}
\centering{\epsfig{figure=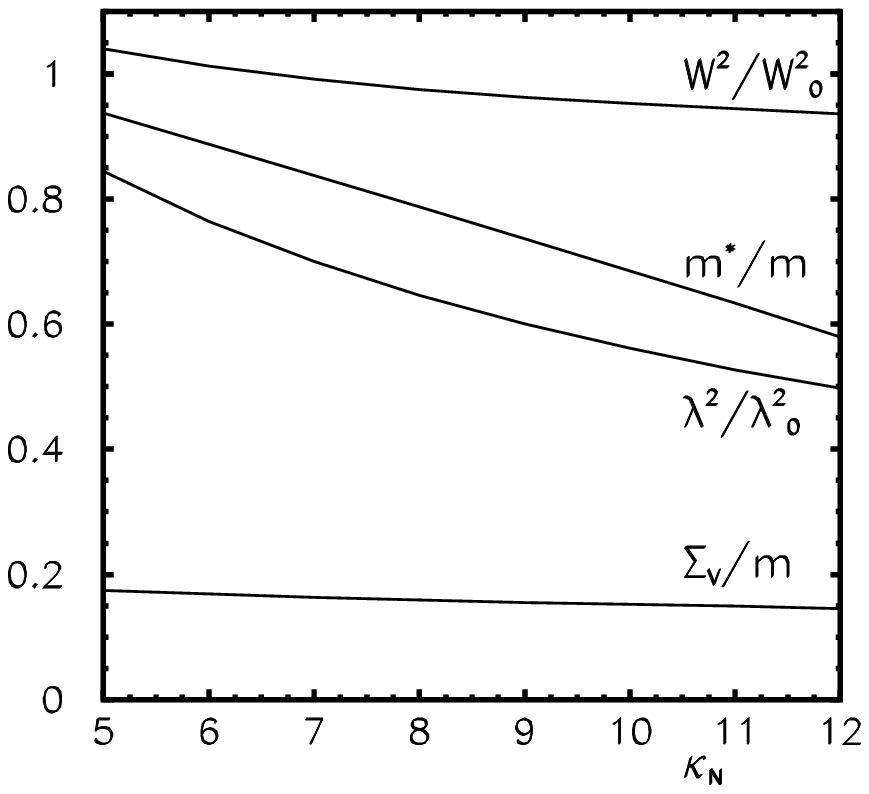,width=10cm}}
\caption{}
\end{figure}

\begin{figure}
\centering{\epsfig{figure=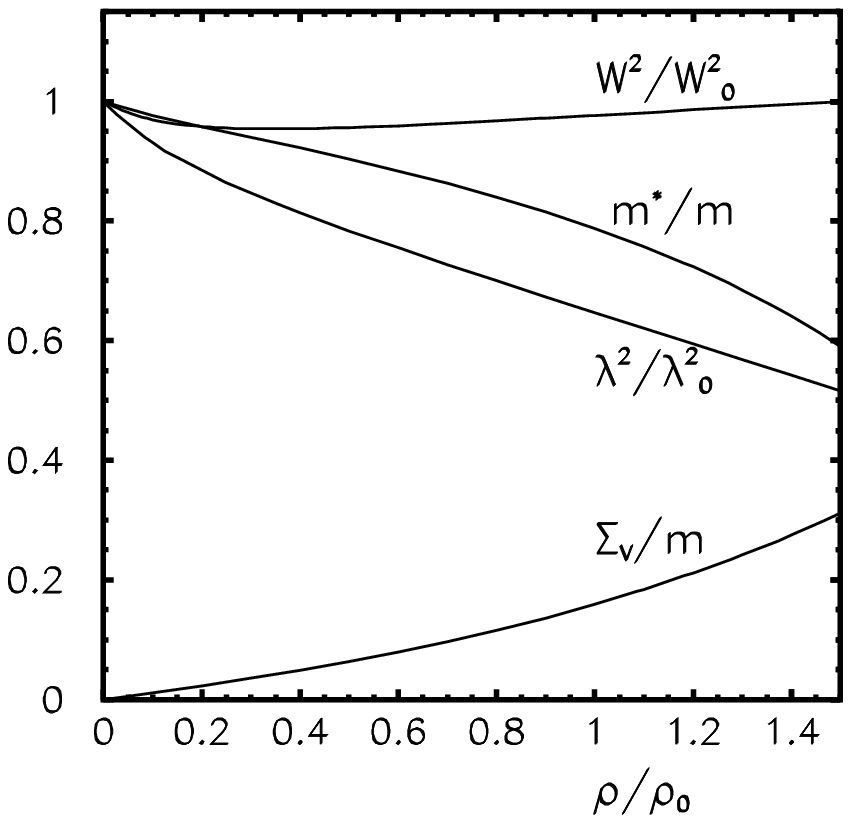,width=10cm}}
\caption{}
\end{figure}

\begin{figure}
\centering{\epsfig{figure=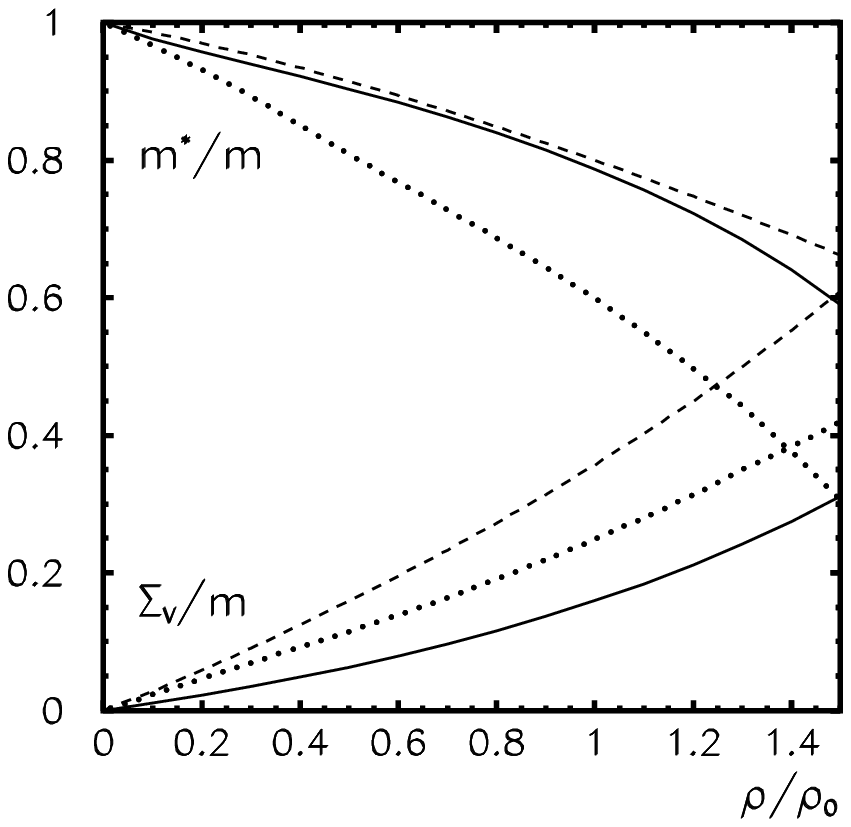,width=10cm}}
\caption{}
\end{figure}

\end{document}